\documentclass[pre,twocolumn,floatfix]{revtex4}
\usepackage{euscript}
\usepackage{bm}
\usepackage{amsmath}
\usepackage{graphicx,mathptmx}
\usepackage{psfrag}
\usepackage{algorithmic}
\def\x{{\bm{x}}}
\def\rn{{\bm{r}}}
\def\ri{{\bm{q}}}
\def\psubi{{\bm{p_q}}}
\def\pn{{\bm{p}}}
\def\pp{{\bm{p}}}
\def\rr{{\bm{r}}}
\def\Rs{{\bm{R}}}
\def\Ps{{\bm{P}}}
\def\F{{\bm{F}}}
\def\I{{\bm{I}}}
\def\xid{\xi^{\dagger}}
\def\mud{\mu^{\dagger}}
\def\H{{\cal{H}}}
\def\Lid{{\cal{L}}^\dagger}
\def\Li{{\cal{L}}}
\def\P{{\cal{P}}}
\def\Pb{{\overline{\cal{P}}}}
\def\Q{{\EuScript{Q}}}
\def\Qb{{\overline{\EuScript{Q}}}}
\def\Qd{{\EuScript{Q}}^\dagger}
\def\Pd{{\cal{P}}^\dagger}
\DeclareMathOperator{\Om}{\Omega}
\DeclareMathOperator{\Omd}{\Omega^\dagger}

\begin{document}

\title{Derivation of a Markov state model of the dynamics of a protein-like chain immersed in an implicit solvent}

\author{Jeremy Schofield}
\email{jmschofi@chem.utoronto.ca}

\author{Hanif Bayat}\email{hbayat@chem.utoronto.ca}

\affiliation{Chemical Physics Theory Group, Department of Chemistry, University of Toronto, \\Toronto, Ontario M5S 3H6, Canada
}

\date{\today}
\begin{abstract}%
A Markov state model of the dynamics of a
protein-like chain immersed in an implicit hard sphere solvent is derived from first principles for a system of   
monomers that interact via discontinuous
potentials designed to account for local structure and bonding in a coarse-grained sense.
 The model is based on the assumption that the implicit solvent interacts on a fast time scale with
the monomers of the chain compared to the time scale for structural
rearrangements of the chain 
and provides sufficient friction so that the motion of monomers is governed by the
Smoluchowski equation.  A microscopic theory for the dynamics of the system is developed
that reduces to a Markovian model of the kinetics
under well-defined conditions. Microscopic expressions for the rate constants that
appear in the Markov state model are analyzed and expressed in terms of a temperature-dependent
linear combination of escape rates that themselves are independent of temperature.
Excellent agreement is demonstrated between the theoretical predictions of the escape rates and
those obtained through simulation of a stochastic model of the dynamics of bond formation.
Finally, the Markov model is studied by analyzing the eigenvalues and eigenvectors of the
matrix of transition rates, and the
equilibration process for a simple helix-forming system from an ensemble of initially 
extended configurations to mainly folded configurations is
investigated as a function of temperature for a number of different
chain lengths. For short chains, the relaxation is primarily single-exponential 
and becomes independent of temperature in the low-temperature regime.  The profile 
is more complicated for longer chains, where multi-exponential relaxation 
behavior is seen at intermediate temperatures followed by a low temperature regime 
in which the folding becomes rapid and single exponential.  It is demonstrated that
the behavior of the equilibration profile as the temperature is
lowered can be understood in terms of the number of relaxation modes or ``folding
pathways'' that contribute to the evolution of the state
populations.
\end{abstract}
\maketitle
\section{Introduction}

In small proteins the competition
between energy (or enthalpy) and
entropy (density of conformational states) leads to transitions between compact
structures, which tend to maximize hydrogen bonding interactions,
and collapsed random coil structures characterized by moderate hydrogen bonding and
increased backbone flexibility.  These competing interactions can also give
rise to a wealth
of other structural transitions among meta-stable states, leading to a free energy landscape that is irregular
and pitted with local minima.  A question of great interest in biological systems is
the connection between features of the free energy landscape such as its roughness, connectivity and 
temperature dependence and their influence on the kinetic profile of the relaxation of a
non-equilibrium ensemble to structures to an equilibrium ensemble.
Such a process is often difficult
to study in atomic detail from first-principles for several
reasons. Perhaps foremost among these is
the computational cost of generating dynamical
trajectories in which large structural
changes can be observed.  

One of the simplest means of extending the time range of a simulation 
is to construct simplified dynamical models of the system.
By simplifying the nature of interactions of complex systems, 
a qualitative view of the mechanisms by which structural transformations
occur can be obtained.  
Under certain conditions, a
simplified picture of the dynamics of the transitions can be formulated by applying ideas
of reaction rate theory in which transitions are treated as first order kinetic
process with well-defined rate constants.  A building block of this approach is to partition the 
free energy landscape into basins corresponding to metastable conformations among which transitions are rare and slow\cite{WalesBook,Noe07,Hummer08}.
In general terms, a metastable state is considered to be
composed of a set of conformations corresponding to a local region of
the free energy surface within which the system evolves on a fast time scale
before crossing a boundary region of higher free energy to another stable region.  If transitions
among locally stable regions are slow compared to that of the motion within the region, then
the dynamics of the overall system may be simplified by ignoring the rapid motion within
the metastable states.  When transitions between states are Markovian, the dynamics of the switching
process between conformations can be described with a Master equation parameterized by a set
of transition rates between connected states.

A major challenge in constructing such a Markov state model\cite{Karpen93,Prinz11,Chodera07,Hummer08,Pan08}
is the identification of the metastable regions of the free energy surface that partition the system
into discrete states.  The identification of metastable states is frequently done by performing 
actual simulations to harvest
trajectories that are then used to cluster conformations of similar geometries, either manually\cite{Caves98,Chodera06}, 
using clustering methods\cite{Karpen93,Bowman09,Keller10,Prinz11a}, or using distance metrics\cite{Karpen93,Chodera07,Bowman09}.   
Rates between states are then estimated based on transitions
observed in trajectories based on discrete state count matrices\cite{Prinz11a}.

In this article a Markov state model of the dynamics of a simple helix-forming protein-like chain 
in an effective solvent in which the constituent beads 
of the chain interact
through discontinuous step potentials is derived from first principles.  
The simple form of the interaction potential permits the 
partitioning of conformational space into states using distance-dependent bonding interactions,
effectively allowing the identification of long-lived states in a temperature-independent fashion without 
resorting to performing any actual molecular dynamics.  It is shown that when transitions between states are slow compared to 
the time scale of motions of the beads in the protein-like chain, 
the transition matrix of rates between states can be computed in terms of
temperature-independent, one-dimensional integrals of distance-dependent densities and
cumulative distributions.  Analytic forms for the densities and cumulative distributions are computed
using fitting procedures on data produced by Monte-Carlo methods.  
Using this apparatus, it is shown how the relaxation profile of 
non-equilibrium populations of states can be computed at arbitrary temperatures.  

The outline of the paper is as follows:  The model for the protein-like chain is introduced in Sec. II, along
with a discussion of the thermodynamic features of the discontinuous potential system.  In Sec. III, expressions
for transitions between configurations are obtained for a chain immersed in an effective bath by analyzing
microscopic expressions for transition rates in terms of the spectral decomposition of the 
projected Smoluchowski evolution operator.  This approach is shown to be equivalent to the first-passage time solution
of the system in a two-step reaction model.  The validity of the solution is demonstrated by direct simulation
of a one-dimensional system.  The rates are incorporated into a Markov state model in Sec. IV and the 
temperature dependent relaxation profile of an ensemble of unfolded configurations is obtained over a range of
effective temperatures.  Conclusions are discussed in Sec. V.

\section{Model}
In this article we analyze the dynamics of a coarse-grained protein model that utilizes step potentials to
characterize interactions between long-lived structural features
of the protein.  Structural elements are enforced using distance constraints in the form of
infinite step potentials while key weak interactions between different regions of the protein are incorporated
through finite step potentials that allow bonds to be formed and broken. Although not considered here, 
important qualitative effects such as misfolding and structural
frustration can be included in this fashion.  The construction of the model
is similar in philosophy to elastic network model
of proteins\cite{Tirion96}, but with harmonic interactions replaced by distance constraints.
 
The use of discontinuous potentials to model interactions in biomolecular systems has been described in detail
elsewhere\cite{Hall2001a,Dokholyan2012}.  In contrast to the simple model introduced below, detailed models
incorporating full all-atom force fields have been developed to account for the real chemical complexity present
in biological systems.  These force fields are
based on underlying continuous potential
models that incorporate both short-ranged local interactions that depend on atom type\cite{Ding2008} as well as long-ranged charge-charge interactions\cite{Dokholyan2012}.  These models have been used to study the folding kinetics of 
small fast-folding proteins\cite{Dokholyan2012} as well as amyloid aggregation and oligomer formation\cite{Hall2004a,Hall2004b,Stanley2006}.
Although there are many bonding interactions in these detailed models, the models are qualitatively similar to the crude
system considered here in that all interactions are in the form of step potentials. In spite of the fact that
the Markov state model derived here applies equally well to all force fields of this form, the construction and analysis of 
a Markov state model  of the dynamics is complicated by the fact that
the number of states in the Markov model may scale exponentially
with the number of bonding interactions in the potential.  
In reality many of the bonding interactions lead to the stabilization
of relatively long-lived minor structural features at temperatures that are relevant to biology.  If the structural
features are long-lived in the temperature range of interest and form quickly in the folding process, the finite step potentials may be coarse-grained and replaced by infinite step potentials, reducing the number of relevant states in
the Markov model.  In principle, such simplifications should arise naturally from the structure of
the transition rate matrix in the Markov model .

As a first step towards examining the influence of the topology of the free energy surface on the 
dynamics of biomolecular systems 
and to explore how the qualitative dynamics change as the roughness of the surface changes with temperature, 
we introduce a simple
discontinuous potential model with relatively few bonding interactions and exhibiting little frustration.  
The model considered here consists of a set of step potentials between beads 
designed to mimic the interactions that lead to the formation of an alpha helix capable of folding back
upon itself. Such interactions have been shown\cite{BayatVanZonSchofield12} to lead to a smooth free energy surface
for short to intermediate length chains of 25 monomers or less.  It should be emphasized that the particular
system analyzed here is not intended to mimic any particular protein, nor does the model represent a general interaction
scheme that can be extended to specific protein systems on the basis of monomer sequence,
but rather to demonstrate how qualitative connections between
the topology of the free energy surface and the dynamics of the system may be analyzed using a Markov state model. 

As a prototypical system,
we consider the dynamics of a beads on a string model\cite{BayatVanZonSchofield12} of a protein-like 
chain in which each bead represents an amino acid or residue. The chain consists of a repeated 
sequence of four different kinds of beads. While having four different types of beads is not 
enough to represent the twenty different types of amino acids, it preserves at 
least some of the qualitative differences between amino acids.

\begin{figure}[ht] 
  \centering
  (a)~~~~~~~~~~~~~~~~~~~~~~~~~~~~~~~~~~~~~~~(b) \\
  \includegraphics[width=0.45\columnwidth]{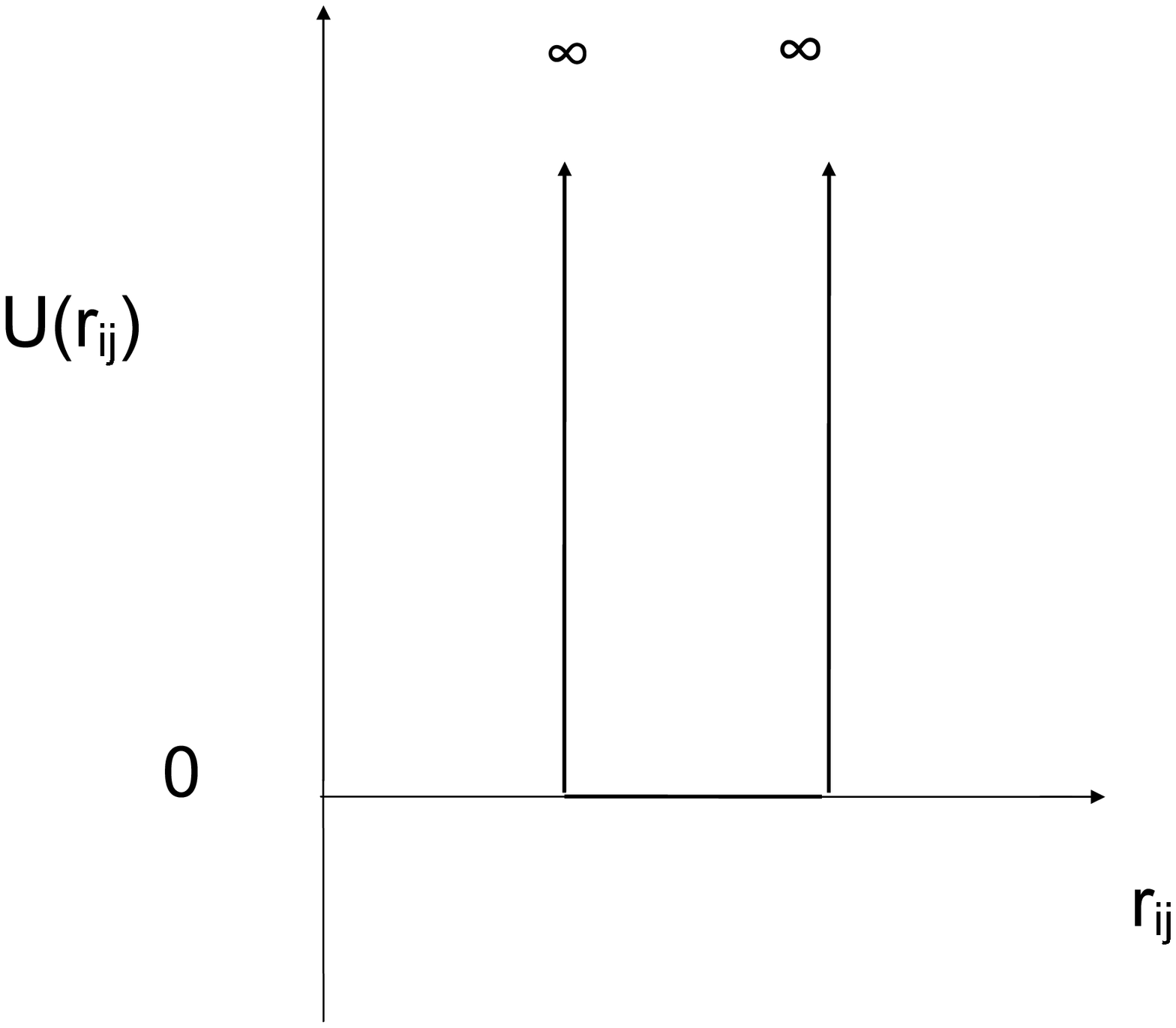}
  \includegraphics[width=0.45\columnwidth]{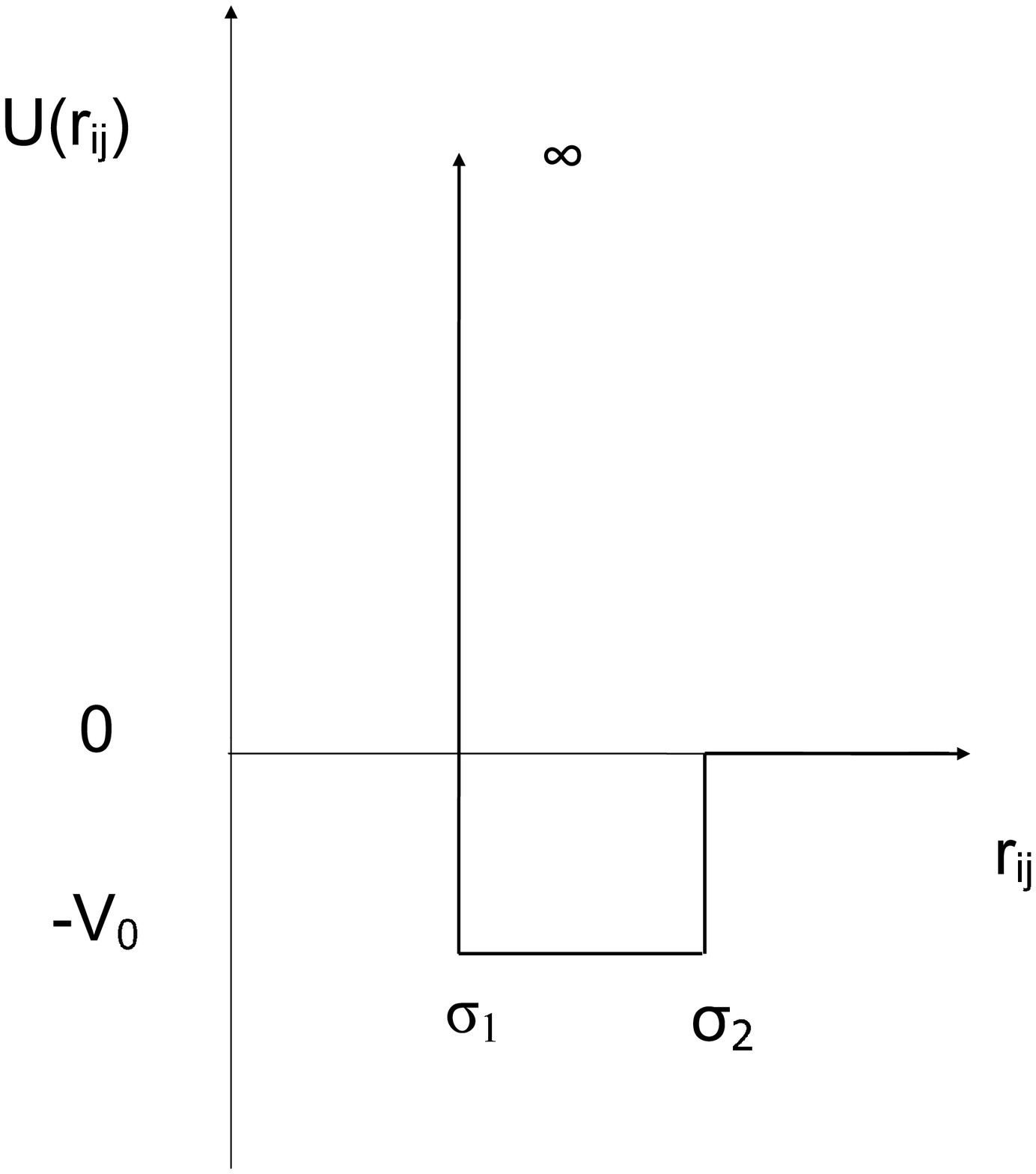}\\
  (c)~~~~~~~~~~~~~~~~~~~~~~~~~~~~~~~~~~~~~~~(d) \\
  \includegraphics[width=0.45\columnwidth]{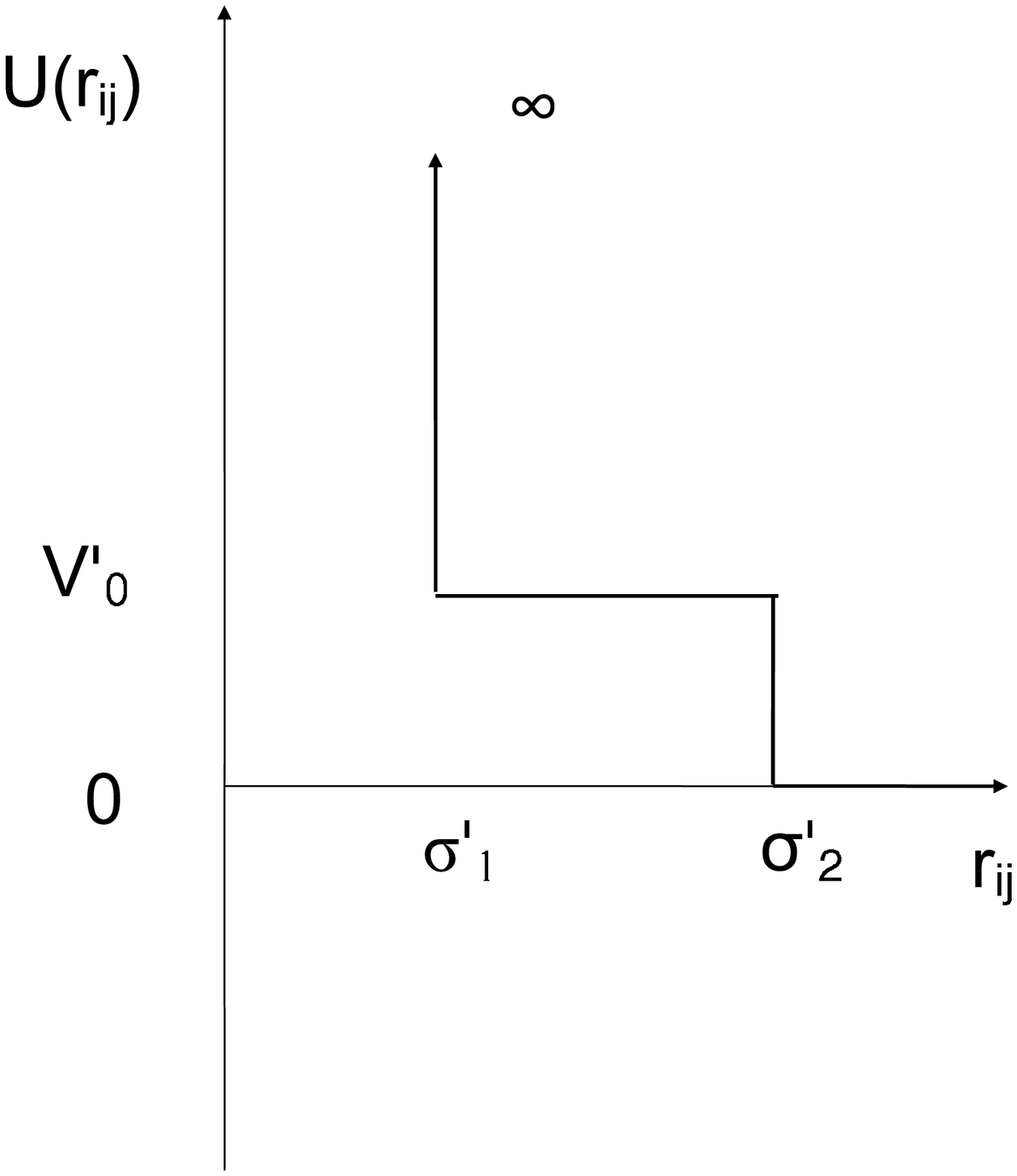}
  \includegraphics[width=0.45\columnwidth]{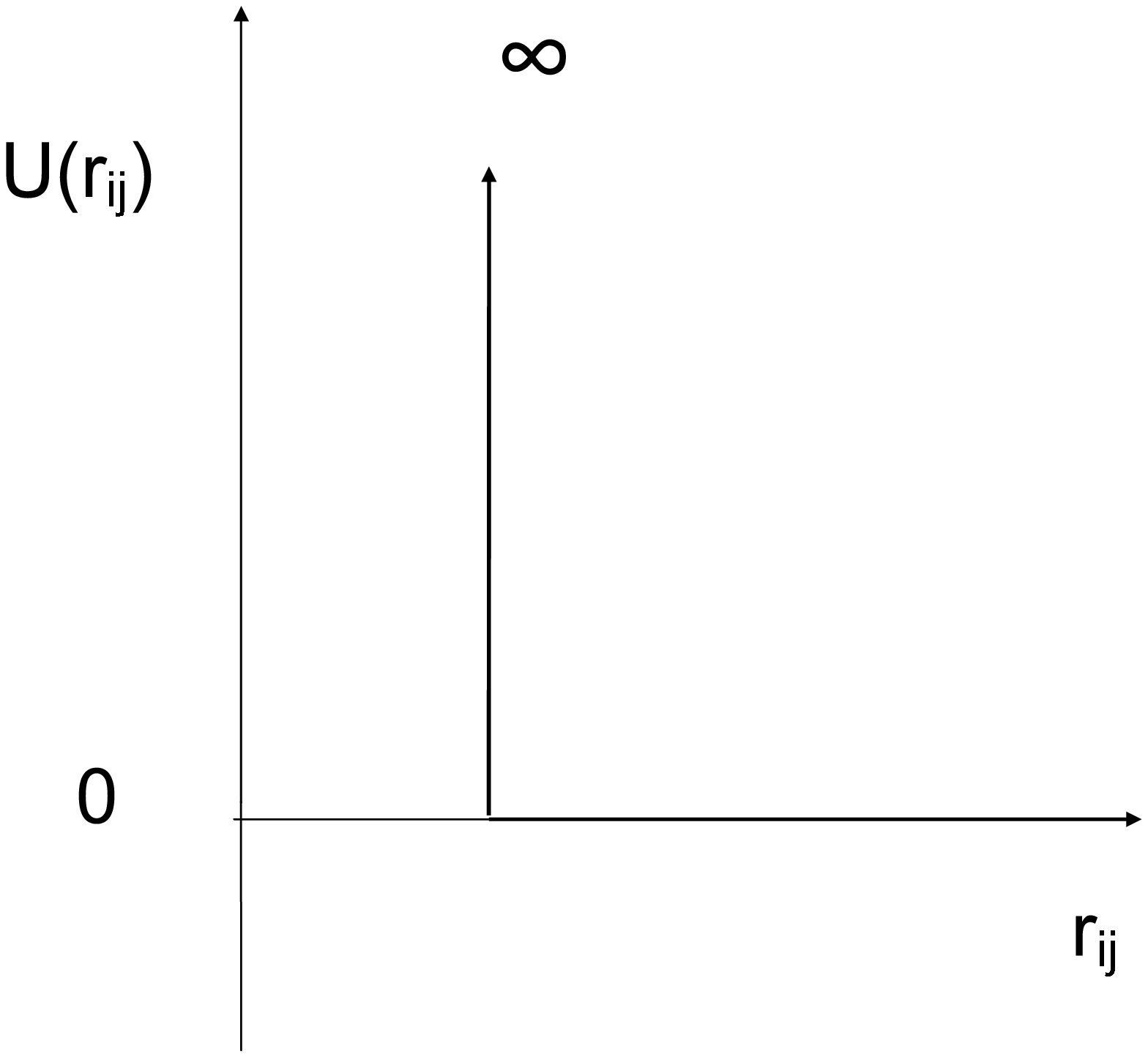}
  \caption{Model~potentials: the (a) infinite square-well potential,
    (b) attractive step potential, (c) repulsive shoulder potentials,
    and (d) hard core repulsion.}
  \label{fig:modelpot}
\end{figure}

In total, four different potentials are used in the model.   The form of these interactions suggests 
physical units to characterize the system.
The first kind of interaction potential acts between the nearest 
and the next nearest neighbors beads in the chain
sequence and restricts the distance between the beads to specific ranges by applying an infinite square-well potential 
similar to Bellemans' bonds model\cite{Bellemans:26} (see Fig.~\ref{fig:modelpot}(a) ). 
To mimic a covalent bond between two consecutive bead elements in the protein, 
the distance between two neighboring beads is restricted to the range $[1.0, 1.17]$ in length units
$\sigma_m$, 
where $\sigma_m$ corresponds to the minimum distance between two adjacent beads.  In Ref.~\onlinecite{BayatVanZonSchofield12}, 
$\sigma_{m}$ was set to the value of $\sigma_{m} = 3.84$~\AA~to mimic the distance between residues in 
real protein systems. 
The infinite square-well potential allows distances between beads to fluctuate around values close to the 
distance between stereocenters used in Ref.~\onlinecite{ZHOU:15}. 
Bond angle vibrations are similarly represented by defining infinite
square-well potentials between next-nearest neighbors in the chain. 
Restricting their distance to a range from $1.4$ to $1.67$ 
generates a vibration angle between 75$^\circ$ and 112$^\circ$. For simplicity, dihedral angle potentials
are not considered in this model,  but as discussed later, some restrictions on non-adjacent beads 
are employed to create rigidity in the backbone of the protein-like chain similar to the 
rigidity that results from the dihedral angle interactions in more detailed potentials.

Attractions between monomers
are modeled by an attractive square-well potential whose depth, $\epsilon$, defines the unit of energy
in the model.  This potential is depicted in Fig.~\ref{fig:modelpot}(b). 
Attractive forces are defined between 
monomers of index $i=4k+2$, and $j=i+4n$, where $k$ is any positive integer (including zero) and $n$ is
any nonzero positive integer excluding $n=2$ and $n=3$. The exclusion of bonding interactions between
monomers separated by $8$ or $12$ beads inhibits the chain from folding back on itself over short distances
and provides an effective chain stiffness that is important for the formation of helical structures\cite{BayatVanZonSchofield12}.
The parameters for the attractive square-well potential, 
$\sigma_1$ and $\sigma_2$, are chosen to be $1.25$ and $1.5$ in the simulation length unit, 
with a mid point that is close to the translation along each turn of an alpha helix 
(5.4~\AA\ in physical units) . These attractive interactions act across longer distances than those in
the covalent interactions.

To represent electrostatic repulsive interactions between monomers, 
a repulsive shoulder potential, shown in Fig.~\ref{fig:modelpot}(c), acts 
between beads $1+4k$ and $4k'$, where $k$ and $k'$ are integers and $k \neq k'$. 
The range of the shoulder is set to be from $1.2$ to $1.9$, 
while the height is $\epsilon$.
The effect of changing the number of step repulsions does not have much
impact on the shape of free energy landscape nor on the dynamics 
of configurations around the native structure. Since the repulsion between the beads increases the 
potential energy while decreasing the configurational entropy, the most common 
structures at low temperatures do not have any repulsive interactions.

\begin{figure}[bth] 
  \centering
  \includegraphics[width=0.5\columnwidth]{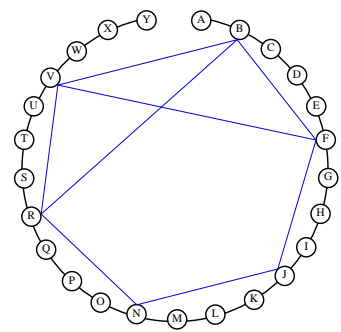}
  \caption{Possible attractive bonds for a chain of 25 beads.}
  \label{fig:interactions}
\end{figure}

Finally, all other bead pairs with no covalent bond, attractive or repulsive interactions 
 interact via a hard sphere repulsion to account for excluded volume interactions at short distances, 
 depicted in Fig.~\ref{fig:modelpot}(d). The hard sphere diameter is set to be $1.25$.

The reduced temperature is defined as $T^*=(k_BT)/\epsilon$, where $\epsilon$ is the potential depth of the attractive square-well interactions, and $\beta^*$ is the inverse of the reduced temperature, $\beta^*=1 /T^*$. 
The unit of time follows from the definitions of the system length, mass and energy scales 
so that $\tau_s = \sqrt{m \sigma^2/\epsilon}$.

\begin{figure}[bth] %
  \centering
  \includegraphics[width=0.8\columnwidth]{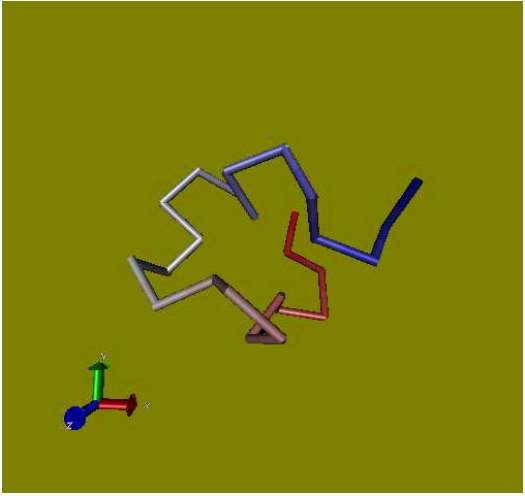}
  \caption{Folded helical structure with 8 bonds.}
  \label{fig:foldedHelix}
\end{figure}

The interactions in the model under consideration here are designed to stabilize compact
helical configurations of the chain at low temperatures\cite{BayatVanZonSchofield12}, 
such as that shown in Fig.~(\ref{fig:foldedHelix}).    In the absence of a 
solvent that can form stable bonds with the chain, 
the system adopts compact configurations at low temperatures
in which all attractive interactions are satisfied.  

The discrete nature of the interactions
allows configurational space to be partitioned into states
by defining an index function for a configuration $c$ that depends on the set of spatial
coordinates $\rn = \{ \rr_1, \dots, \rr_N \}$ of the chain with $N$ monomers,
\begin{eqnarray*}
\chi_c( \rn ) = \left\{ 
\begin{array}{ll}
1 & \mbox{if only bonds in $c$ are present,} \\
0 & \mbox{otherwise}.
\end{array}
\right.
\end{eqnarray*}
The partitioning of configurational space arises naturally by
expanding the product in the identity
\begin{eqnarray}
1 &=& \prod_{i=1}^{n_b} \left( 1 - H(x_i - x_c) + H(x_i-x_c)\right) \nonumber \\
&=&
\prod_{i=1}^{n_b} 
\left( H^b(x_i - x_c) + H(x_i-x_c)\right) \\
 &=& \sum_{k=1}^{n_s} \chi_{c_k} (\rn) , 
\label{partition}
\end{eqnarray}
where $n_b$ is the number of attractive bonds in the model, $n_s =
2^{n_b}$ is the number of states, $H^b(x) = 1 - H(x)$, 
and $H(x)$ is the Heaviside function
\begin{eqnarray*}
H( x ) = \left\{ 
\begin{array}{ll}
1 & x \geq 0 \\
0 & \mbox{otherwise}.
\end{array}
\right.
\end{eqnarray*}
In Eq.~(\ref{partition}), $x_i$ is the distance between monomers in
the $i$th bond, and $x_c$ is the critical distance at which a bond is
formed.  For notational simplicity, we order the index of
configurations based on the number of bonds starting with the
configuration with no bonds, $\chi_1(\rn ) = \prod_{i=1}^{n_b} H (x_i
- x_c)$, and ending with the configuration with the maximum number of
bonds, $\chi_{n_s}(\rn ) = \prod_{i=1}^b H^b(x_i-x_c)$.

Using the index functions, the entropy $S_c$ of a configuration at an
inverse temperature $\beta = 1/(k_B T)$ can be
computed as
\begin{eqnarray*}
S_c = \frac{3}{2} N k_B \ln \left( \frac{2\pi m e}{\beta h^2} \right)
+ k_B \ln  \frac{1}{V^N} \int^\prime d\!{\rn} \; \chi_c (\rn), 
\end{eqnarray*}
and the relative entropy of two configurations $c_1$ and $c_2$ is
\begin{eqnarray*}
\Delta S_{c_1 c_2} =  k_B \ln \frac{\int^\prime d\!\rn \; \chi_{c_1}
  (\rn)}{\int^\prime d\!\rn \; \chi_{c_2} (\rn} = \frac{\Delta U_{c_1
    c_2}}{T}
+ k_B \ln \frac{f_{\mbox{obs}}(c_1, T)}{f_{\mbox{obs}}(c_2, T)},
\end{eqnarray*}
where the integral over Cartesian positions $\rn = \{ \rr_i \}$ is restricted to 
configurations that satisfy all
geometric constraints due to the infinite square-well and hard core repulsions.
Here $\Delta U_{ij}$ is the potential energy difference between
configurations $i$ and $j$ and $f_{\text{obs}}(i, T)$ is the probability
of observing configuration $i$ at temperature $T$.  This latter
quantity is estimated numerically using Monte Carlo methods. 
Due to the nature of the interaction potentials, the entropic differences between
configurations is independent of temperature and can be estimated
from equilibrium simulations based on the relative populations
observed.  From these simulations, it is evident that the formation of
a bond is accompanied by a loss of configurational entropy typically
on the order of $3 \; k_B$.  

\begin{table}[b] 
  \begin{tabular} {r|l|r|r}
     & configuration & $U_c/\epsilon$ & $S_c/k_B$ \\
    \hline
    1 & No Bond& 0.00 & 31.8 $\pm$0.6 \\
    2 & BF & -1 & 28.6 $\pm$ 0.6\\
    3 & BF JN & -2 & 25.1 $\pm$ 0.6 \\
    4 & BF NR  & -2 & 25.2 $\pm$ 0.6\\
    5 & BF JN RV & -3 & 21.7 $\pm$ 0.4\\
    6 & BF FJ NR RV  & -4 & 17.8 $\pm$ 0.6\\
    7 & BF FJ JN RV  & -4 & 17.6 $\pm$ 0.6\\
    8 & BF FJ JN NR RV & -5 & 13.2 $\pm$ 0.6\\
    9 & BF BR BV FJ JN NR RV & -7 & 3.7 $\pm$ 0.8\\
    10 & BF BV FJ FV JN NR RV & -7 & 2.9 $\pm$ 0.6\\
    11 & BF BR BV FJ FV JN NR RV & -8 & 0
  \end{tabular}
  \caption{Potential energy in units of $\epsilon$ and relative entropy of the common structures of the 25-bead chain.}
  \label{tab:modelB-dominant-entropy}
\end{table}

Since the configurational space can be unambiguously partitioned into
states whose equilibrium populations can be computed, one can
also estimate the cumulative distribution functions, probability
densities, and potential of mean force associated with the formation
of a bond.  For example, consider a chain of $25$ monomers that are
alphabetically indexed so that the first monomer is labeled ``A'',
the second ``B'', and so on (see Fig.~\ref{fig:interactions}).  Since the bonding interactions are
determined by distances between pairs of monomers, bonds and therefore
configurations can be identified by alphabetic pairings, such as $BF BR$, 
which indicates a configuration in which the second monomer is
bound to the sixth and eighteenth monomers and no others.  The
configuration {\it A} = $BF BR$ can be formed from the configuration {\it B} = $BF$ by
the formation of the $BR$ bond, which occurs when the distance
$x_{BR} = |\bm{r}_{B} - \bm{r}_{R}|$ is less than the critical 
bond formation distance $x_c$.  One can define a probability density
$\rho_{a}(x)$  and the cumulative distribution 
$C_{a}(x) = \int_{x_{\rm{min}}}^{x} dy \;
\rho_{a}(y)$ in terms of canonical ensemble averages 
restricted over states
as
\begin{eqnarray*}
\rho_{a}(x) &=& \left\langle \delta(x - x_{BR}) \right\rangle_{a},
\label{probDensity}
\end{eqnarray*}
where the notation $\langle f(\rn) \rangle_{a}$ denotes the
normalized uniform average 
\begin{eqnarray*}
\left\langle f(\rn) \right\rangle_{a} = \frac{\int^\prime d\!\rn
\; \chi_{a}(\rn) \, f(\rn)}{\int^\prime d\!\rn
\; \chi_{a}(\rn)} = \frac{\int^\prime d\!\rn
\; \chi_{a}(\rn) \, f(\rn)}{Z_A}.
\end{eqnarray*} 

The relative entropy, shown in Table~\ref{tab:modelB-dominant-entropy}, and other equilibrium 
properties such as the probability density and cumulative distribution function for bonds can be 
computed by Monte Carlo simulation\cite{BayatVanZonSchofield12}.  From such simulations, 
one simple way to estimate $\rho_{a}(x)$ is to construct histograms
of the distance $x_{BR}$ for configurations that satisfy the bonding constraints
of configuration {\it A}.  Since the
probability density and the cumulative distribution function are
independent of temperature, the distance $x_{BR}$ from 
any instantaneous configuration that
satisfies the bonding criteria for configuration {\it A} can be used.  A
more appealing way of constructing analytic fits to the densities
and cumulative distribution functions is to use a
procedure\cite{VanZonSchofield10} that constructs these quantities
from sampled data using statistical fitting criteria.

The temperature-dependent potential of mean force $\phi_{ab}(x)$ connecting states
{\it A} and {\it B} can be computed from the probability densities
$\rho_{a}(x)$ and $\rho_{b}(x)$  by first considering the
cumulative distribution function connecting the two states,
\begin{eqnarray*}
C_{ab}(x) = \left\{
\begin{array}{ll}
\frac{e^{\Delta S/k_B} e^{\beta \epsilon}}{1 + e^{\Delta S /k_B} e^{\beta
    \epsilon}}
C_{a}(x) &   x_\text{min} \leq x < x_c \\
\frac{1}{1 + e^{\Delta S/k_B} e^{\beta
    \epsilon}} C_{b}(x) + 
\frac{e^{\Delta S/k_B} e^{\beta \epsilon}}{1 + e^{\Delta S/k_B} e^{\beta
    \epsilon}} & x_c \leq x \leq x_\text{max}
\end{array}
\right. ,
\label{cumulativeReaction}
\end{eqnarray*}
where $\Delta U_{a b} = U_a - U_b = -\epsilon$ and $\Delta S = S_{a} - S_{b}$ is the relative
entropy difference of the configurations (see Table~\ref{tab:modelB-dominant-entropy}).  Noting that $\rho_c(x) =
dC_c(x)/dx$, we find that
\begin{eqnarray*}
\rho_{ab}(x) = \left\{
\begin{array}{ll}
\frac{e^{\Delta S/k_B} e^{\beta \epsilon}}{1 + e^{\Delta S/k_B} e^{\beta
    \epsilon}}
\rho_{a}(x) &  x_\text{min} \leq x < x_c \\
\frac{1}{1 + e^{\Delta S/k_B} e^{\beta
    \epsilon}} \rho_{b}(x)  & x_c \leq x \leq x_\text{max}
\end{array}
\right. ,
\label{densityReaction}
\end{eqnarray*}
which is discontinuous due to the nature of the potential at $x=x_c$.
In principle, the densities are also discontinuous and vanish abruptly above and below some maximum and minimum
values $x_{\text{max}}$ and $x_{\text{min}}$ that are determined by the
distance constraints imposed by the infinite square-well interactions present in the model.
The potential of mean force, $\phi_{ab}(x) = -k_B T \ln \rho_{ab}(x)$, 
describes the reversible work associated with
pulling the system from configuration {\it B} to {\it A} by reducing the
distance $x=x_{BR}$ between monomers $B$ and $R$.  Since the density is discontinuous
at $x=x_c$, the potential of mean force has a jump discontinuity of
magnitude 
\begin{eqnarray*}
\Delta \phi_{ab} &=& \phi_a(x_c^-) - \phi_b(x_c^+) \approx -\epsilon,
\end{eqnarray*}
where $\phi_a(x)$ is the potential of mean force in region {\it A} where
$x < x_c$ defined by $\exp (-\beta \phi_a (x) ) / Z_A = \rho_a(x)$,  $\phi_b(x)$ is the potential of mean force
in region {\it B} where $x \geq x_c$,
 and $x_c^{-}$ and $x_c^{+}$ are the left and right-side limits as $x$ approaches $x_c$.
Due to the simplicity
of the interaction potential, the potential of mean force can be
computed at any inverse temperature $\beta$ from the temperature
independent densities $\rho_{a}$ and $\rho_{b}$.  The density
$\rho_{ab}(x)$ can be written as a conditional equilibrium average
of a delta function via
\begin{eqnarray*}
\rho_{ab}(x) = \left\langle \delta (x - x_{BR})
  \chi^k_{b} (\rn ) \right\rangle,
\end{eqnarray*}
where $\chi_{b}^k(\rn )$ is the indicator function for configuration
{\it B} {\it without} a factor of $H(x_{\text{BR}} - x_c)$ or
$H^b(x_{\text{BR}} - x_c)$ indicating whether or not $x_{\text{BR}} \geq x_c$.
Note that the potential of mean force $\phi_{ab}(x)$
includes an effective volume factor of the form $-2\beta^{-1} \ln x$ 
that arises from the conversion of Cartesian into spherical polar
coordinates.  This volume factor leads to a temperature independent 
entropic barrier for the formation of a bond.

\begin{figure}
\includegraphics[width=0.6\columnwidth]{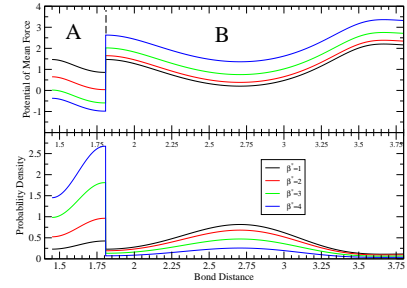}
\caption{The potential of mean force (top panel) and the probability density (lower panel) in dimensionless units as a
  function of the dimensionless critical bond distance. The regions defining the bound state {\it A} and unbound state
  {\it B} are identified.}
\label{reactionPMF}
\end{figure}

In Fig.~(\ref{reactionPMF}), the potential of mean force is plotted as a
function of the $NR$ distance for configurations {\it B} = $BF BR BV F\!J J\!N
RV$ and {\it A} = $BF BR BV F\!J J\!N NR RV$ for a range of different inverse temperatures.
Note the entropic barrier between the non-bonded states with $x \geq x_c = 1.8$
and bonded states with $x < x_c$.  The overall shape of the potential
of mean force in the stable wells in regions {\it A} and {\it B} is non-trivial due to the
complicated geometric constraints imposed by the infinite square well potentials.
The temperature independent form of the potential must therefore be computed 
numerically for all but the simplest of systems.
We shall see in the next section that the densities $\rho_i (x)$ and
cumulative distributions $C_i (x)$ play an important role in determining
the rate constants relating the rate of population transfer between states.

\section{Dynamics}
The system considered here consists of $N$ monomers of mass $m$ of a protein-like chain immersed in
a solvent of light particles of mass $m_s$.  The monomers in the model are assumed to correspond to a coarse-grained
representation of either amino acid residues or even sequences of residues, so that the difference in
masses between the monomers and solvent leads to a separation of time scale of monomer and solvent motions.
Under such conditions, one finds that to second order in the small parameter $\sqrt{m_s/m}$, the
evolution of the probability density $f(\rn, \pn, t)$ is governed by the Fokker-Planck equation\cite{Oppenheim70},
\begin{eqnarray}
\frac{\partial f(\rn,\pn,t)}{\partial t}
 &=& \Om f(\rn, \pn ,t ),
\label{FPevolution}
\end{eqnarray}
with formal solution $f(\rn ,\pn ,t) = \exp\{\Om t\} f(\rn, \pn, 0)$,
where the Fokker-Planck operator is given by
\begin{eqnarray}
\Om &=& \sum_{i=1}^{N} \left[ -\frac{\pp_i}{m} \cdot \nabla_{\rr_i} + \nabla_{\rr_i} U  \cdot \nabla_{\pp_i}
+ \gamma \nabla_{\pp_i} \cdot \left( \nabla_{\pp_i} + \beta \frac{\pp_i}{m} \right) \right] \nonumber \\
&=&  -\frac{\pn}{m} \cdot \nabla_{\rn} + \nabla_{\rn} U  \cdot \nabla_{\pn}
+ \gamma \nabla_{\pn} \cdot \left( \nabla_{\pn} + \beta \frac{\pn}{m} \right) ,
\label{Smoluchowski}
\end{eqnarray}
where $\gamma$ is the friction coefficient and $U$ is the potential of mean force describing monomer-monomer
interactions averaged over the bath.  In Eq.~(\ref{Smoluchowski}), 
we have defined the dot product over monomer vector positions and momenta as
$\rn \cdot \rn = \sum_{i=1}^N \bm{r}_i \cdot \bm{r}_i$, where
$\bm{r}_i$ is the position vector of monomer $i$. 
We assume the interaction potential of mean force $U$ can be written as a sum of pairwise interactions that depend only
on the relative distance between monomers, and hence can be expressed as
\begin{eqnarray}
U &=& \sum_{i=1}^N \sum_{j=i+1}^N u(r_{ij}) \nonumber \\
&=& u(r_{12}) + \sum_{i=3}^N \big( u(r_{i1} + u(r_{i2}) \big) + \sum_{i=3}^N 
\sum_{j=i+1}^N u(r_{ij}) \nonumber \\
&=& U_0 (r_{12}) + \phi (\rn ) + U_{i} (\rn),
\label{potDecomposition}
\end{eqnarray}
where $U_0$ is the direct interaction potential for a selected pair of monomers $1$ and $2$, $\phi$ is the interaction 
potential of the selected pair with all other monomers in the chain, and $U_i$ is the interaction potential
for all other monomers.
For the purpose of discussion below, we assume that the
square-well and hard-core potentials have been smoothed out to continuous potentials in which the
potential energy increases sharply but continuously over short distances.

For many model systems, the bonding interactions due to interaction terms of the form 
$u(r_{ij}) = H(r_{ij} - r_c) - \epsilon H^b(r_{ij} - r_c)$ involve pairs of monomers
that are separated by many other monomers in the chain.  One expects that the mean interaction time $\tau_c$ for such
bonding terms is much larger than the time scale $\tau_l$ of local interactions of neighboring monomers
due to the connectivity of the chain.  Hence it is reasonable to expect that the time scale of
the transitions between configurations, which is proportional to $\tau_c$, are slow compared to the
time scale of the motion of the monomers $\tau_l$, so that the dynamics of populations $\bm{c}(t) = \{
c_1(t), \dots , c_n(t) \}$ of the states
is well-represented by a simple Markov model for $t > \tau_l$:
\begin{eqnarray}
\dot{\bm{c}}(t) = \bm{K} \cdot \bm{c}(t),
\label{MarkovModel}
\end{eqnarray}
where $\bm{K}$ is an $n_s \times n_s$ matrix of transition rates
connecting the $n_s$ states. In this dynamical picture, an arbitrary non-equilibrium density equilibrates 
on a time scale $\tau_l$ to a local equilibrium form in which states are distributed with non-equilibrium
populations.

In the following section, Eq.~(\ref{MarkovModel}) is justified from first principles
and microscopic expressions for the transition rates are obtained.

\subsection{Microscopic expressions of the transition matrix}
It is possible to express the transition rates in the matrix $\bm{K}$
in terms of time-dependent correlation functions using standard 
methods of statistical mechanics.  To see how this is done, we first note that the
population $c_i(t)$  of state $i$ can be obtained from the
evolution of the non-equilibrium average of the indicator function $\chi_i (\rn)$ since
\begin{eqnarray*}
c_i (t) &=& \int d\rn d\pn \; \chi_i (\rn ) f(\rn, \pn, t) \nonumber \\
 &=&  \int d\rn d\pn \; \chi_i(\rn) e^{\Om t} f(\rn, \pn, 0 ) ,
\end{eqnarray*}
where $f(\rn ,\pn ,0)$ is the initial non-equilibrium distribution.  For
simplicity, we will assume that the initial density is of local equilibrium form
$f(\rn ,\pn, 0) = Z^{-1} \sum_i p_i \chi_i (\rn
) e^{-\beta \H}$, where the $p_i \geq 0$ are constants dependent
on the initial population of the states and $Z$ is the
normalization constant
\begin{eqnarray*}
Z = \sum_i p_i \; \int d\rn d\pn \; \chi_i(\rn ) e^{-\beta \H} =
\sum_i p_i e^{-\beta F_i} , 
\end{eqnarray*}
where $F_i = -n_i \epsilon + T S_i$ is the free energy of
state $i$ with $n_i$ bonds of energy $-\epsilon$ and 
entropy $S_i$.  This initial
distribution corresponds to an ensemble of states in which the
distribution within a given state is of equilibrium form even
though the actual populations $c_i$ of the states are not at their
equilibrium values.  The form assumed for the initial non-equilibrium distribution 
is not critical in the subsequent analysis,
and presumably the local equilibrium form applies to any
non-equilibrium initial state for times $t$ satisfying $\tau_c \gg t > \tau_l$.

To obtain microscopic expressions for the transition rates, one
defines a projection operator $\P$ and its complement $\Q = \bm{I} -
\P$ that act on an arbitrary density $g(\rn ,\pn)$ as
\begin{eqnarray*}
\P g(\rn ,\pn ) &=& \sum_{i=1}^{n_s} \frac{\chi_i (\rn) \rho_e }{
\left\langle  \chi_i (\rn ) \right\rangle
} \int d\rn d\pn \; \chi_i (\rn ) g(\rn ,\pn) , 
\label{Projection}
\end{eqnarray*}
where $\left\langle B(\rn ,\pn) \right\rangle = \int d\rn d\pn \; \rho_{e}(\rn, \pn )
B(\rn ,\pn)$ denotes the equilibrium average over the stationary
distribution of the Fokker-Planck equation $\Om \rho_e(\rn ,\pn ) = 0$. Here
the equilibrium density is
\begin{eqnarray*}
\rho_e (\rn ,\pn ) &=& \frac{e^{-\beta \H}}{\int d\rn d\pn \; e^{-\beta \H}}.
\end{eqnarray*}

Applying standard operator identities to the Fokker-Planck equation~(\ref{FPevolution}),
we obtain\cite{Kapral89},
\begin{eqnarray}
c_{i}(t) - c_i (0) = \sum_{j=1}^{n_s} \int_0^t d\tau \; K_{ij}(\tau) \; c_{j}(t-\tau),
\label{nonMarkov}
\end{eqnarray}
where $K_{ij}(t) = k_{ij}(t)/\left\langle \chi_j \right\rangle$ and
\begin{eqnarray*}
k_{ij}(t) &=& \int d\rn d\pn \; \chi_i \Om e^{\Q \Om t} \bigg( \chi_j \rho_e \bigg).
\label{GammaDef}
\end{eqnarray*}

If there is a separation of time scale between the time scale $\tau_c$ of
the evolution of the
populations $c_i (t)$ and the time scale $\tau_l$ over which the
correlation
function $K_{ij}(t)$ decays to its constant plateau value,
Eq.~(\ref{nonMarkov}) can be approximated by a Markovian form for $t \gg \tau_l$
\begin{eqnarray}
c_{i}(t) - c_i (0)
&=& \sum_{j=1}^{n_s} K_{ij} \int_{0}^t d\tau \; c_{j}(\tau) + O(\tau_l/\tau_c ),
\label{Markov}
\end{eqnarray}
where $K_{ij} (t) \rightarrow K_{ij}$ for $t \gg
\tau_l$.  From Eq.~(\ref{Markov}), we observe that the populations obey the equation of motion
\begin{eqnarray*}
\dot{c}_{i}(t) = \sum_{j=1}^{n_s} K_{ij} c_{j}(t),
\end{eqnarray*}
as in Eq.~(\ref{MarkovModel}), and indicates the populations exhibit exponential (or finite
multi-exponential) behavior.

It is straightforward to show that 
$\Om (\rho_e B) = \rho_e \Omd B$ and $\Q \Om (\rho_e B) = \rho_e \Qd \Omd B$, where $\Omd$ and
$\Qd$ are the Hermitian conjugates of the operators $\Om$ and $\Q$, respectively, given by
\begin{eqnarray*}
\Omd &=& \frac{\pn}{m}\cdot \nabla_\rn - \nabla_\rn U \cdot \nabla_\pn + \gamma \left( \nabla_\pn - \beta 
\frac{\pn}{m} \right) \cdot \nabla_\pn \nonumber \\
\Pd A(\rn, \pn) &=& \sum_{i=1}^{n_s} \frac{\left\langle A \chi_i \right\rangle}{\left\langle \chi_i \right\rangle} \chi_i
\nonumber \\
\Qd &=& 1-\Pd \nonumber . 
\end{eqnarray*}
Using these results, we find that
\begin{eqnarray}
k_{ij}(t) &=& \left\langle \chi_i \left( e^{\Omd \Qd t} \Omd \chi_j 
  \right) \right\rangle = 
 \left\langle \left( \Omd e^{\Qd \Omd t}
    \chi_i \right) \chi_j 
  \right\rangle \nonumber \\
  &=& \left\langle \left( e^{\Omd \Qd t}
    \Omd \chi_i \right) \chi_j
  \right\rangle .
\label{NonMarkovRates}
\end{eqnarray}
Noting  $\sum_{i}^{n_s} \chi_i = 1$, it is evident from Eq.~(\ref{NonMarkovRates}) 
that the diagonal terms
can be written in terms of the off-diagonal elements 
$K_{jj}(t) = - \sum_{i \neq j} K_{ij}(t)$.
The final expression in Eq.~(\ref{NonMarkovRates}) can be interpreted
as the evolution of the initial flux $\Omd {\chi_i}$ sampled from
configuration $j$ under the projected operator $\Omd \Qd$. Using Eq.~(\ref{NonMarkovRates}), we
see that the transition matrix elements satisfy
\begin{eqnarray*}
K_{ij}(t) 
  &=& K_{ij}(t)
\frac{\langle \chi_i \rangle}{\langle \chi_j \rangle}
\nonumber \\
&=& K_{ji}(t) \, e^{-\beta (F_i - F_j)},
\label{detailedBalance}
\end{eqnarray*}
which establishes that the rates in the transition matrix obey the condition of detailed balance.  

If $j > i$,
the initial state $j$ has more bonds than state $i$, 
indicating that the dynamics has led to a barrier crossing out of at least one bonding
well.  We associate the rate of
barrier crossing out of a bonding state with a forward rate and interpret
$K_{ij} =k^f_{ij}$ when $j > i$.
On the other hand if $i > j$, then the
initial configuration $j$ has fewer bonds than configuration $i$,
and at least one bonding barrier has been crossed from the reverse direction,
so that $K_{ij} = k^r_{ij}$.  Note that in both cases we
expect $K_{ij} > 0$ since the population of state $i$ increases.
We therefore interpret the rates as
\begin{eqnarray*}
\lim_{t \rightarrow \infty} K_{ij}(t) = \left\{ 
\begin{array}{ll}
k^r_{ij}  & i >  j \\
k^f_{ij} & j > i,
\end{array}
\right.
\end{eqnarray*}
where $k^f$ and $k^r$ denote forward and reverse rate
constants for the formation and loss of the bond distinguishing state $i$
from state $j$.

The rate expressions can be regarded as an average of the dynamical
variable $K_{i}(\rn , \pn,t) = e^{\Omd \Qd t} \Omd \chi_i
(\rn )$ over the conditional equilibrium density of configuration
$j$, $\rho_e (\rn, \pn) \chi_j (\rn )$.  Note that $K_{i}(\rn ,\pn ,t)$ obeys the evolution equation,
\begin{eqnarray}
\frac{\partial K_{i}(\rn, \pn, t)}{\partial t} = (\Omd \Qd)
K_{i}(\rn,\pn, t).
\label{MemoryEvolution}
\end{eqnarray}
If the Markov
description of the dynamics is sensible, the elements of the matrix
$K_{ij}(t)$ rapidly approach their asymptotic values after a short transient time
$\tau_m \sim \tau_l$.  The transient time $\tau_m$ can be estimated by examining
the eigenvalue spectrum of the projected evolution operator 
$\Omd \Qd$, and corresponds roughly with
the inverse of the smallest nonzero eigenvalue $\mu^\dagger_{\text min}$.

The dynamical variable $K_{i}(\rn , \pn,t)$ depends on all $6N$ phase space coordinates $(\rn ,\pn )$ of the
protein-like chain, which complicates its spectral analysis.  
In the following section, we show that this dependence can be approximately reduced to 
a single coordinate in the high friction limit provided there is a clear separation of time scale between events leading to bond formation or destruction.  

\subsection{Evaluation of rate constants}
The computation
of an element  $K_{ab} (t)$ of the rate matrix that characterizes the rate of 
transitions between two states  {\it A} and {\it B} can be interpreted as
the average of $K_{a}(\rn, \pn, t)$ over an equilibrium distribution of bond distances 
in the {\it B} configuration,
\begin{eqnarray}
K_{ab} (t) &=& \frac{\left\langle \big[  K_{a} (\rn, \pn , t) \big]
\, \chi_b \right\rangle}{\langle \chi_{b} \rangle} 
= \frac{k_{ab}(t)}{\langle \chi_{b} \rangle} .
\label{fullKab}
\end{eqnarray}
Recall that from Eq.~(\ref{MemoryEvolution}), 
the time-dependence of $K_a(\rn ,\pn ,t)$ is governed by the projected evolution operator
$\Omd \Qd$.
If the transient
time scale $\tau_m$ is small compared to the reactive time scale $\tau_c$,
it is reasonable to assume that only a single barrier separating configurations
{\it A} and {\it B} is crossed for any given trajectory and hence
non-zero transition rates $K_{ab}$ only occur
between states that differ by a {\it single}
bond dependent on the scalar distance $x = r_{mn}$
between monomers $m$ and $n$.
Under these conditions, it is shown in Appendix A that the transition rate matrix 
$k_{ab}(t)$ may be written as a one-dimensional integral over the conditional equilibrium probability
density $\rho_{ab}(x)$ defined in Eq.~(\ref{densityReaction}) as
\begin{equation}
k_{ab}(t) =  \int_{x_\text{min}}^{x_\text{max}} dx \; \rho_{ab}(x) 
\, H(x - x_c)  \, e^{\Lid_{xq} t} \Lid_x H(x-x_c) . 
\label{k_alphabeta1}
\end{equation}
In Eq.~(\ref{k_alphabeta1}), the evolution operators $\Lid_x$ and $\Lid_{xq}$ are given by
\begin{eqnarray}
\Lid_x g(x) &=&  D \left( \frac{d^2}{dx^2} - \beta {\phi}_{ab}^\prime (x) \frac{d}{dx} \right) g(x) \nonumber \\
\Lid_{xq} g(x) &=& \Lid_x \Qd_x g(x) 
\label{evolutionApprox}
\end{eqnarray}
where the prime notation indicates the derivative, the effective diffusion coefficient $D$ is
$D = 2 (k_B T)^2 / \tilde{\gamma}$, where $\tilde{\gamma}$ is the total friction on the beads (defined in the Appendix), 
and $\phi_{ab}(x)$ is the potential of mean force defined by $\phi_{ab}(x) = - k_B T \, \log\rho_{ab}(x)$. 
In Eq.~(\ref{evolutionApprox}), the projection operator $\Qd_x$ removes the projection of a density $g(x)$
onto the conditional equilibrium density $\rho_{ab}(x)$ so that
\begin{eqnarray}
\Qd_x g(x)&=& (1-\Pd_x) g(x) \nonumber \\
&=& g(x) - \rho_a(x) H^b(x-x_c) \int_{x_{\rm{min}}}^{x_c} dx \; g(x) \nonumber \\
&&- \rho_b(x) H(x-x_c) \int^{x_{\rm{max}}}_{x_c} dx \; g(x).
\end{eqnarray}

In the above discussion, we have defined state {\it A} to be the bonded state with conditional probability
density $\rho_A(x) = H^b(x-x_c) \rho_{ab}(x)/\langle H^b(x-x_c) \rangle$, where the brackets
$\langle B(x) \rangle = \int dx \; \rho_{ab}(x) B(x)$, whereas the unbonded state {\it B} has a conditional probability density $\rho_B(x) = H(x-x_c) \rho_{ab}(x)/\langle H(x-x_c) \rangle$.  Subsequently, we drop subscripts
for quantities like the density $\rho(x) = \rho_{ab}(x)$ and the potential of 
mean force $\phi (x) = \phi_{ab}(x)$ with the understanding that
the we are concerned with the rate constants $K_{ab}$ for transitions between states {\it A} and {\it B}.

\subsection{Diagonalization approach to rate constants}
The direct evaluation of the rate constants using a spectral
decomposition of the evolution 
operator $\Lid_{xq}$ 
is feasible due to its low dimensionality.  One
difficulty to overcome is the discontinuity in the potential of
mean force $\phi (x)$ at the positions
$x_{\text{min}}$ and $x_{\text{max}}$ arising from the infinite repulsive potentials, 
as well as the jump discontinuity at $x=x_c$.  The infinite repulsions lead
to reflecting boundary conditions at $x=x_{\text{min}}$ and $x=x_{\text{max}}$, while
the discontinuity at $x=x_c$ leads to jump conditions at the singularity, as discussed in Appendix B.

In principle,
solutions of the Smoluchowski equation with unprojected dynamics can
be considered as well.  The solution of the Smoluchowski equation for regions between
discontinuities must be considered
separately and then patched together by using the appropriate boundary
jump conditions.  Exact solutions of the one-dimensional Smoluchowski equation
have been considered using this approach for square-well
potentials\cite{Morita94,Morita96} and for square-wells with 
square barriers\cite{Felderhof08}.  A valuable aspect of the
analytical treatment of such problems is the allowance of a comparison
of the perturbative expansion of the solution with approximate
solution methods, such as the first passage time approach.
For one-dimensional systems with square-wells and square barriers, it is possible to find
corrections to Kramer's solution for the survival probability to find a particle in a 
meta-stable well\cite{Kramers40}.  When the
barriers separating meta-stable states are not sufficiently high, the 
probability to find a particle in a well exhibits 
a multiple exponential relaxation profile\cite{Felderhof08}.

The evolution of the system under the projected Smoluchowski operator
$\Lid_{xq}$ is more difficult to analyze than the unprojected operator $\Lid_x$\cite{Kapral89}. 
Nonetheless, the
eigenfunctions $\{ \xid_n (x) \}$ and the corresponding real eigenvalues
$\{ \mud_n \}$ of the operator $\Lid_{xq}$ can be found 
even when the potential of mean force exhibits jump discontinuities.

At this point, it is useful to define the inner product between
real functions $f^\dagger(x)$ and $g(x)$ as
\begin{eqnarray*}
\langle f | g \rangle = \int_{x_\text{min}}^{x_\text{max}} dx \; f^\dagger(x) g(x).
\end{eqnarray*}
The adjoint of an operator $\hat{O}$ is formally defined as
\begin{eqnarray*}
\langle f| \hat{O}g \rangle &=& \int_{x_\text{min}}^{x_\text{max}} dx \; f^\dagger(x) \, \hat{O}g(x) \\
&=& \int_{x_\text{min}}^{x_\text{max}} dx \; \left( \hat{O}^\dagger f^\dagger(x) \right) \, g(x) \\
&=& \langle \hat{O}^\dagger f | g \rangle . 
\end{eqnarray*}

The eigenvalue equations for $\Lid_{xq}$ and its adjoint $\Li_{xq} = \Q_x \Li_x$,
\begin{eqnarray}
\left( \Lid_{xq}  + \mu^\dagger_n \right) \xi^\dagger_n(x) &=& 0 \nonumber \\
\left( \Li_{xq} + \mu_n \right) \xi_n(x) &=& 0 ,
\label{eigenvalueEqn}
\end{eqnarray}
are second order homogeneous linear differential equations of the
Sturm-Liouville form.  
It follows from the Sturm-Liouville form of Eqs.~(\ref{eigenvalueEqn}) that the left and right eigenvectors 
$\xid_n(x)$ and $\xi_n(x)$ 
are orthonormal\cite{Kapral89}
\begin{eqnarray*}
\int_{x_\text{min}}^{x_\text{max}} dx \; \xid_n(x) \xi_m(x) = \langle \xi_n | \xi_m
\rangle = \delta_{n,m},
\end{eqnarray*}
and constitute a complete set so that $F(x) = \sum_{n=0}^{\infty} \langle
\xi_n | F \rangle \, \xi_n(x)$ for an arbitrary function $F(x)$ defined on
the interval $[x_\text{min}, x_\text{max}]$.

Returning to the computation of rate constants, we note that inserting complete
sets of eigenvectors in the expression for the $k_{ab}$ in Eq.~(\ref{k_alphabeta1}) for $a \neq b$ gives
\begin{eqnarray*}
k_{ab} &=& - \left\langle  e^{\Lid_x \Qd_x t}\Lid_x H(x-x_c) \big| \rho H(x-x_c)  \right\rangle \nonumber\\
&=& -
\sum_n \langle H(x-x_c) | \Li_x \xi_n \rangle \langle \xi_n | e^{\Li_{xq} t} | \xi_n \rangle \langle \xi_n | H(x-x_c) \rho \rangle \nonumber \\ 
&=&  -\sum_{n=0} a_n b_n e^{-\mu_n t} ,
\end{eqnarray*}
where we have defined expansion coefficients 
$a_n = \langle  H(x-x_c) | \Li_x \xi_n \rangle$
and $b_n = \langle \xi_n |H(x-x_c) \rho (x) \rangle$.  
Integrating by parts, the expansion coefficients $a_n$ can be expressed as
\begin{eqnarray*}
a_n &=& - D \int_{x_\text{min}}^{x_\text{max}} dx \; 
H^\prime(x-x_c) \left(
  \xi^\prime_n(x) +\beta \phi^\prime(x) \xi_n(x) \right) \nonumber \\
&=& - D \left( \xi_n^\prime(x_c) + \beta \phi^\prime (x_c) \xi_n(x_c)
\right) ,
\end{eqnarray*}
where the reflecting boundary conditions at $x_\text{min}$ and $x_\text{max}$ have been applied.
As shown in Appendix B, the expansion coefficients $a_n$ are well-defined in
the presence of jump discontinuities due to the continuity of the probability current, 
as is evident from Eq.~(\ref{jump2}).  Analyzing the spectrum of the projected operator
$\Li_{xq}$ (see Appendix B), we find there are two eigenvectors $\xi_0$ and $\xi_1$ with zero eigenvalue.
However since the equilibrium density is stationary, we find that $a_0 = 0$ so that only the second eigenvector
$\xi_1$ contributes to the long time limit of the rate constant, which approaches the constant value
\begin{eqnarray}
k_{ab} &=& -a_0 b_0  -a_1 b_1 = - a_1 b_1,
\label{kab}
\end{eqnarray}
at long times.

In Appendix B an analytic expression for the long time limit of $k_{ab}$ is obtained 
that allows the inverse rate coefficients to be computed from integrals of the densities $\rho_a(x)$ and $\rho_b(x)$ and 
cumulative distributions $C_a(x)$ and $C_b (x)$ using the relation
\begin{eqnarray}
D \; k^{-1}_f 
&=& \int_{x_{\text{min}}}^{x_c} dx \; \frac{C_a(x)^2}{\rho_a(x)} \nonumber \\
&& \; \; \; + 
e^{\beta \epsilon} e^ {\Delta S / k_B}  \int_{x_c}^{x_{\text{max}}} dx \;
\frac{(1-C_b(x))^2}{\rho_b(x)}  \nonumber \\
D \; k^{-1}_r 
&=&  e^{-\beta \epsilon} e^ {-\Delta S / k_B} 
\int_{x_{\text{min}}}^{x_c} dx \; \frac{C_a(x)^2}{\rho_a(x)}  \nonumber \\
&& \; \; \; +  \int_{x_c}^{x_{\text{max}}} dx \;
\frac{(1-C_b(x))^2}{\rho_b(x)} .
 \label{finalForwardBackward}
\end{eqnarray}

Remarkably, these rate expressions can be expressed in terms of the
average first passage time between the states since the average first passage
time out of state {\it A} into state {\it B} is given by\cite{Schulten80,Schulten81}
\begin{eqnarray}
\langle \tau (a) \rangle &=& 
D^{-1} \int_{x_\text{min}}^{x_c} dy \,
\frac{C_a(y)^2}{\rho_a(y)} ,
\label{fpa}
\end{eqnarray}
whereas the average first passage time out of state B, is
\begin{eqnarray}
\langle \tau (b) \rangle  &=& D^{-1} \int_{x_c}^{x_{\text{max}}} dy \, \frac{ \left( 1 -
    C_b(y) \right)^2}{\rho_b(y)}. 
\label{fpb}
\end{eqnarray}
These results lead to simple forms for the inverse rate constants
\begin{eqnarray*}
k^{-1}_f &=& \langle \tau (a) \rangle + 
e^{\beta \epsilon} e^ {\Delta S / k_B}  \langle \tau (b) \rangle  \nonumber \\
k^{-1}_r &=&  e^{-\beta \epsilon} e^ {-\Delta S / k_B} \langle \tau (a) \rangle
 +  \langle \tau (b) \rangle .
\nonumber
\end{eqnarray*}

In Eq.~(\ref{finalForwardBackward}), we note that the ratio of
the population of state {\it A} to the population of state {\it B} is $Z_A/Z_B =
e^{\beta \epsilon} e^ {\Delta S / k_B}$, where the bond depth
$\epsilon > 0$ and the entropic difference $\Delta S < 0$.
For low temperatures where $\beta$ is large, note that $k_r^{-1}$ is
dominated by the second term and becomes independent of
temperature. The forward rate constant, $k_f$, becomes increasingly
smaller at low temperatures since the free energy barrier to escape
out of the bonding well increases with temperature.  Due to the
simplicity of the model, the cumulative distribution functions $C(x)$
and densities $\rho (x)$ are independent of temperature for states {\it A}
and {\it B}, and hence the temperature dependence of the forward and reverse
rate constants is determined by the ratio of the relative equilibrium populations
of the two states.

The expressions for the forward and reverse rate constants obtained through the
spectral decomposition of the projected Smoluchowski operator are equivalent to
those obtained from a two-step reaction in a simple 3-state model
\begin{eqnarray*}
A
\begin{array}{c}
{\mbox{$k_1$ }}\\
{\mbox{\Large $\rightleftharpoons$}} \\[-2mm]
{\mbox{$k_{-1}$}}
\end{array}
C
\begin{array}{c}
{\mbox{$k_2$ }}\\
{\mbox{\Large $\rightleftharpoons$}} \\[-2mm]
{\mbox{$k_{-2}$}}
\end{array}
B ,
\end{eqnarray*}
where state C is defined to be the region near the dividing surface $x=x_c$.
  Writing
first-order mass action 
kinetic equations for the time evolution of the populations $N_A(t)$, $N_B(t)$ and $N_C(t)$
for the two step reaction and assuming
the steady state approximation $dN_C/dt = 0$ to eliminate two of the
rate constants, we find the effective rate equations
\begin{eqnarray*}
\frac{dN_A}{dt} &=& - k_f \, N_A + k_r \, N_B \\
\frac{dN_B}{dt} &=&  -k_r \, N_B + k_f \, N_A,
\end{eqnarray*}
where 
\begin{eqnarray}
k_f^{-1} &=& k_1^{-1} + k_{-2}^{-1} \frac{Z_A}{Z_B} \nonumber \\
k_r^{-1} &=& k_{-2}^{-1} + k_{1}^{-1} \frac{Z_B}{Z_A} .
\label{newRates}
\end{eqnarray}
To obtain Eq.~(\ref{newRates}), we have used the detailed
balance condition, $k_f/k_r = Z_B/Z_A = (k_1 k_2)/(k_{-1}k_{-2})$.  In equilibrium, the relative
populations of A and B are $Z_A/(Z_A + Z_B)$ and $Z_B/(Z_A + Z_B)$,
respectively, and the relaxation for a system initially in state B
obeys
\begin{eqnarray}
N_B(t) = \frac{Z_B}{Z_A + Z_B} + \left( 1 - \frac{Z_B}{Z_A+Z_B}
\right) e^{-(k_f + k_r) t},
\label{populationDynamics}
\end{eqnarray}
with characteristic relaxation time $(k_f + k_r)^{-1}$.  If the
rate constants $k_1$ and $k_{-2}$  
are approximated by the inverse first passage time out of the stable wells to an absorbing state
at $x=x_c$, namely we set $k_1 = 1/\langle \tau (a) \rangle$ and $k_{-2} = 1/\langle \tau (b) \rangle$, 
Eqs.~(\ref{newRates}) and (\ref{finalForwardBackward}) coincide.

The forward
and reverse rate constants can be computed without
much difficulty from the analytic fits to the cumulative
distribution functions $C(x)$ and reduced densities $\rho (x)$ for
states {\it A} and {\it B}.
In addition to obtaining expressions for the rate constants, the
spectrum of the projected evolution operator can be computed by expanding the densities, cumulative
distributions, and potentials of mean force in a basis set.  It is
then possible to evaluate the contribution of eigenvectors with
non-zero eigenvalues to the time dependent correlation functions
$K(t)$ to estimate the transient time scale $\tau_m$.

\subsection{Numerical test of microscopic rate expressions}
In principle the quality of the description outlined in the previous
section can be examined by computing the spectrum of the projected
evolution operator and evaluation of the transient time $\tau_m$.  If
there is an insufficient separation of time scale so that $\tau_m$ is
similar in magnitude to the typical time $\tau_c$ required to cross
barriers, then the population dynamics obtained in
Eq.~(\ref{nonMarkov}) is inherently non-Markovian.  One expects that
the separation of time scale is sensitive to the barrier height
between states and possibly the shape of the potential of mean
force as well.  The calculation of $\tau_m$ is a relatively heavy
task, involving the evaluation of matrix elements $\langle l | \Lid \Q
| m \rangle$ in some complete set of basis functions $|l \rangle$.

A much more direct and simple way to verify that an adequate
separation of time scale holds is to simulate the Smoluchowski
dynamics of the populations
under the appropriate potential of mean force and verify that the
population dynamics show exponential decay with decay time $(k_f +
k_r)^{-1}$ predicted by Eq.~(\ref{finalForwardBackward}) , as
suggested in Eq.~(\ref{populationDynamics}).

\begin{figure}[htb]
\includegraphics[width=0.6\columnwidth]{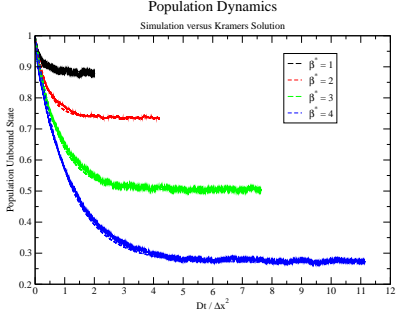}
\caption{The equilibration dynamics of the population of the unbound
  state based on an ensemble size of $10^4$ random walkers.  
The solid lines correspond to simulation results, and the
  dashed lines are analytical fits based on the Kramer's first-passage
  time solution for the potential of mean force in
  Fig.~\ref{reactionPMF}.  The different curves correspond to
  different effective temperatures $\beta^* = \beta \epsilon$.  Note
  that at the lowest effective temperature $\beta^* = 4$, the
  equilibrium bound
  population is significantly larger than the unbound population.}
\label{mcSim}
\end{figure}

To this end, an initial non-equilbrium system in which all members of an
ensemble evolve from an initial state of conditional equilibrium in
state {\it B} according to the effective potential in
Fig.~(\ref{reactionPMF}) was simulated.  The simulation was done using
a Monte Carlo procedure in which steps of magnitude $\pm \Delta x$
were attempted with equal probability and accepted with probability
$\text{min}(1, \, \rho(x_t)/\rho(x))$, where $x$ is the current state of
the system, $x_t = x \pm \Delta x$ is the trial configuration, and
$\rho (x)$ is the probability density in
Eq.~(\ref{densityReaction}). The simulation was done under conditions
with diffusion coefficient $D = (\Delta x)^2/\Delta t = 1$, so that
the discrete system time evolved with $\Delta t = 1/\Delta x^2$.  For
the simulation results shown in Fig.~(\ref{mcSim}), $\Delta x = 0.01$.
As is clear from the results shown in Fig.~(\ref{mcSim}), the decay of
the population of the unbound state is roughly single-exponential and well described by the
theoretically predicted rate for the range of temperatures examined.
Small deviations from single exponential decay are evident in the equilibration dynamics at short times,
particularly when $\beta^*$ is small.  Deviations from single-exponential relaxation indicate
the influence of other non-dominant eigenmodes to the relaxation profile.
The integrals in Eqs.~(\ref{fpa}) and (\ref{fpb}) for the intermediate 
rate constants $k_1$ and $k_{-2}$
were carried out numerically using analytical fits to the cumulative
distributions and densities, and it was found that $k_1^{-1} =
0.0287$, and $k_{-2}^{-1} = 1.532$.
\subsection{Smoluchowski dynamics of a simple model}
The rate expressions can be validated for a simple model system consisting
of two bound particles interacting via step potentials 
immersed in an implicit solvent.  In the model the interaction between
the particles in the binary system depends only on the distance $r$
between particles and is given by
\begin{eqnarray*}
U(r) = \left\{
\begin{array}{ll}
-\epsilon & \mbox{$r < r_c$} \\
0 & \mbox{$r_c \leq r < R$} \\
\infty & \mbox{$r \geq R$}
\end{array}
\right. .
\end{eqnarray*}

In the model system, the dynamics
of the system consists of evolution of the bound particles under the interaction potential
for a fixed period of time $\tau_{\text{rot}}$ followed by a collision step\cite{MalevanetsKapral99,MalevanetsKapral00}
in which the effect of solvent particle collisions with the particles in 
the system is introduced through a collision operator that modifies
the particle velocities $\bm{v}_i$ according to a collision rule\cite{SchofieldInderKapral12,GIKW09,kikuchi02,kikuchi03}
\begin{eqnarray}
\bm{v}_i^\prime = {\bm{v}}_c + \bm{\omega}_k \cdot (\bm{v}_i - {\bm{v}}_c),
\label{collisionRule}
\end{eqnarray}
where $\bm{v}_i^\prime$ is the post-collision velocity of particle $i$ and $\bm{\omega}_k$ is a randomly chosen
rotation matrix.
In Eq.~(\ref{collisionRule}), ${\bm{v}}_c$ is a local velocity defined by
\begin{equation*}
\bm{v}_c = \frac{m}{m_T} \bm{v}_i + \frac{n_s m_s}{m_T} \bm{v}_s,
\end{equation*}
where $m_s$ is the mass of a solvent particle and 
$n_s$ is drawn from a Poisson distribution with mean value
$\rho$, where $\rho$ is the number density of solvent in the system, here taken to be $10$
in simulation units. Here the total mass is $m_T = m + n_s m_s$, and $\bm{v}_s$ 
is an effective solvent velocity drawn from a Maxwell-Boltzmann distribution 
with mass $n_s m$.  Since this velocity is drawn at each collision step, the 
velocity of the solvent is uncorrelated from one collision step to another.  
Since each particle behaves as a point particle, the only contribution to the self-diffusion coefficient comes from the rotation collision step.  Hence for this model, the decay of the velocity autocorrelation function for an isolated bead is single exponential and the self-diffusion coefficient for each particle is given by\cite{SchofieldInderKapral12}
\begin{equation*}
D =  \frac{\tau_{\text{rot}}}{\beta^* \xi} \left( \frac{\epsilon}{m \sigma^2} \right)^{1/2} = 
\left(\frac{\tau_{\text{rot}}}{\tau_s} \right) \frac{1}{\beta^* \xi}
\end{equation*}
where $\tau_s$ is the time unit of the system, and $\xi = 2 \gamma_s/(2-\gamma_s)$ is the effective solvent friction and
\begin{equation*}
\gamma_s =  (1- c_\gamma ) e^{-\rho} \sum_{n=1}^{\infty}
\frac{\rho^n}{n!} \frac{n}{\mu + n},
\end{equation*}
where $c_{\gamma}$ is the average diagonal element of the randomly chosen rotation matrices, 
$c_{\gamma} = \sum_k \text{Tr} \, \bm{\omega}_k /3$, and $\mu = m/m_s$ is the mass ratio of the system and solvent particles, here taken to be $20$.  For the simple model, the average collision frequency in inverse time units is computed
to be
\begin{eqnarray*}
\nu = \left( \frac{9}{\pi \beta^*} \right)^{1/2} \left( \frac{\sigma}{r_c} \right)
\frac{e^{\beta^*} + a^2 + 1}{e^{\beta^*} + a^3 - 1},
\end{eqnarray*}
where $a = R/r_c$. At low temperatures where $\beta^*$ is large, the typical time scale of interaction $\tau_i$ is
approximately $\sqrt{\pi r_c^2 \beta^*}/3$.

\begin{figure}[htb]
\includegraphics[width=0.8\columnwidth]{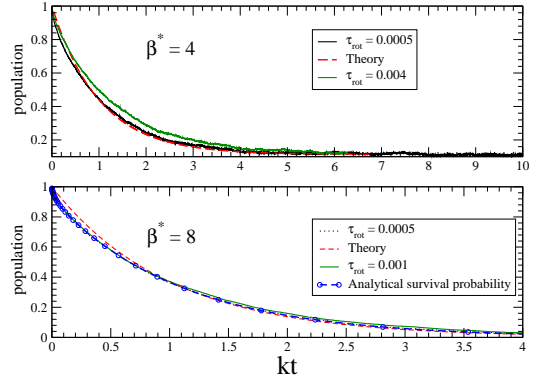}
\caption{The equilibration dynamics of the population of the unbound
  state as a function of scaled time $kt$, where $k = k_r + k_f$. The top panel 
  shows the relaxation profile at $\beta^* = 4$, and the bottom panel shows detail of the profile at $\beta^* = 8$.
  The solid lines correspond to simulation results for different values of the parameter $\tau_{\text{rot}}$, and the
  dashed line is the prediction for the time dependence of the unbound population based on Eq.~(\ref{finalForwardBackward}).  In the botton panel, the dotted line with circles is the analytical solution of the survival probability for a particle diffusing between a reflecting sphere at $R$ and an absorbing sphere at $r_c < R$
   where $R/r_c = 2$.}
\label{srdFigure}
\end{figure}

For $\tau_i \gg \tau_{\text{rot}}$, the dynamics is diffusive and dynamical variables evolve according
to the Smoluchowski operator.  From Eqs.~(\ref{fpa}) and (\ref{fpb}), we find that
\begin{eqnarray*}
\left\langle \tau (a) \right\rangle &=& \frac{r_c^2}{30 D} \\
\left\langle \tau (b) \right\rangle &=& \frac{r_c^2}{30D} \frac{5a^6 - 9a^5 + 5a^3 -1}{a^3 -1} ,
\end{eqnarray*}
and hence
\begin{eqnarray*}
Dk_f^{-1} = \frac{r_c^2}{30} \left( 1 + \frac{e^{\beta^*}}{(a^3 - 1)^2} 
\left( 5a^6 - 9 a^5 + 5 a^3 -1 \right) \right) ,
\end{eqnarray*}
while $k_r^{-1} = (a^3 - 1) e^{-\beta^*} k_f^{-1}$.  As in the previous example, the evolution of 
the population of
an initial ensemble of unbound states with $r > r_c$ was computed numerically 
for a choice of $r_c = \sigma$ and $a = 2$ 
at different dimensionless inverse temperatures $\beta^*$.  As is evident in Fig.~(\ref{srdFigure}), the 
decay of the unbound population to its equilibrium value is well approximated by the exponential decay at 
different values of $\beta^*$.  At $\beta^* = 4$, the rate of decay
$k_f + k_r =0.02507$ predicted by the analytic solution when $\tau_{\text{rot}}/\tau_s = 0.002$
clearly underestimates the decay rate at short times and overestimates the decay at intermediate times.  
Deviations at short times from the single exponential
decay are even more evident at low temperatures, as is clear from the bottom panel of Fig.~(\ref{srdFigure}).
For these parameters, the average frequency of collisions is roughly $0.8$ in dimensionless units, while
the bare self-diffusion coefficient at $\beta^*=4$ and $\tau_{\text{rot}} = 0.001$ is $D=0.0038$, 
more than two orders of magnitude smaller.  
These deviations from simple exponential behavior can be understood by considering the behavior of the model
at low temperatures (large $\beta^*$) where the stability of the bound state and large barrier to unbind
simplify the dynamics.  In the low temperature regime, the decay of the ensemble of unbound states is equivalent
to the decay of the survival probability of a particle diffusing between two concentric spheres of radii $r_c$ and $R$ 
where the boundary conditions for the diffusing equation are reflecting at $R$ and absorbing at $r_c$.
The survival probability $P_s(t)$ is computed to be the inverse Laplace transform of the function
$\tilde{P_s}(z)$, where
\begin{eqnarray*}
\tilde{P_s}(z) &=& \frac{1}{z} \left( 1 + \frac{3 x_{c}^2}{x_R^3 - x_c^3}
\frac{i_1(x_c) k_1(x_R) - k_1(x_c) i_1 (x_R)}{i_0(x_c) k_1(x_R) + k_0(x_c) i_1(x_R)} \right) 
\end{eqnarray*}
where $x_c = r_c \sqrt{z/D}$, $x_R = R \sqrt{z/D} $, and $i_\alpha$ and $k_\alpha$ are modified spherical Bessel functions of the first and second kind of order $\alpha$.  The survival probability has the property that
$\tilde{P_s}(0) = \langle \tau (b) \rangle$, which approaches $k^{-1} \approx k_r^{-1}$ as $\beta^*$ increases. 
This result implies that even though the profile of the decay of the survival probability is non-exponential,
the areas under the curves of the survival probability $P_s(t)$ and the predicted exponential decay are equal.  Hence
the faster initial decay of the survival probability is canceled by the slower decay at later times.
 In  the bottom panel of Fig.~(\ref{srdFigure}), the survival probability obtained by numerically inverting
the Laplace transform of $\tilde{P_s}(z)$ using the Stehfest algorithm\cite{Stehfest70a}
 is plotted versus the scaled time.
It is evident from the plot that the relaxation profile of the simulated system is essentially exact for $\beta^* = 8$.
Furthermore, for $kt \gg 1$, the differences between the survival probability and the exponential decay are small.  
In this sense, the theoretical prediction represents the best single exponential fit to the actual decay 
observed in the system.

The agreement between the theoretically predicted and numerically computed populations is 
less impressive at larger values of $\tau_{\text{rot}}$.  These results suggest that rapid collisions
of the particles with the solvent are necessary to guarantee the conditions of diffusive reaction dynamics,
probably due to the relatively small entropic barrier between the bound and unbound states.

\section{Markov model of the configurational dynamics of the helix-forming chain}
We now consider the evolution of an initial non-equilibrium ensemble of configurations of the
protein-like chain introduced in Sec. II.
Based on the analysis of the test cases in the previous section, it is clear that
under conditions of large friction and reasonably low temperature $\beta^* \geq 1$,
the dynamics of the fractional populations $\bm{c}(t) = \{
c_1(t), \dots , c_n(t) \}$ of the states
is well-represented by a simple Markov model:
\begin{eqnarray}
\dot{\bm{c}}(t) = \bm{K} \cdot \bm{c}(t),
\label{MarkovModel1}
\end{eqnarray}
with formal solution $\bm{c}(t) = e^{\bm{K} t} \bm{c}(0)$,
where $\bm{K}$ is an $n_s \times n_s$ matrix of transition rates
connecting the $n_s$ states of the helix-forming system.  
We recall that since the sum of the
fractional populations is one, $\sum_\alpha
c_\alpha (t) = 1$, the diagonal elements of $\bm{K}$ are given by 
\begin{eqnarray*}
K_{\alpha \alpha} = - \sum_{\beta \neq \alpha} K_{\beta \alpha} ,
\end{eqnarray*}
whereas the off-diagonal rates are
\begin{eqnarray*}
 K_{\alpha \beta} = \left\{ 
\begin{array}{ll}
k^r_{\alpha \beta} & \alpha > \beta \\
k^f_{\alpha \beta} & \beta > \alpha,
\end{array}
\right.
\end{eqnarray*}
where the forward and backward rates between states $\alpha$ and
$\beta$ are given by Eq.~(\ref{finalForwardBackward}).  

Since it is assumed that there is a clear separation of time scale
between reactive events and the transient time scale of the evolution
within a given configuration, the matrix element $K_{\alpha \beta} =
0$ for any two states $\alpha$ and $\beta$ which differ by more than a
single bond.  This property means the $n_s \times n_s$ matrix is
sparse, particularly for systems with a large number of possible
bonds.  For larger systems, the equilibrium population of certain
states with small entropies compared to other configurations with the
same number of bonds is effectively zero over the entire temperature
range, and such states need not be considered.  From the form of the rate
constants, it is expected that the rate of transitions from other states to
these rare states is relatively small and justifies their neglect.

A dynamical picture of the equilibration of an initial non-equilibrium
ensemble of states can be obtained by diagonalization of the matrix
$\bm{K}$ to write
\begin{eqnarray}
\bm{c}_\alpha (t) &=& \bm{Q}_{\alpha \beta} \, e^{-\lambda_\beta t} \,
\bm{Q}^{-1}_{\beta \gamma} \, \bm{c}_{\gamma}(0),
\label{matrixEvolution}
\end{eqnarray}
where summation over repeated Greek indices is implied.  In
Eq.~(\ref{matrixEvolution}), the columns of the transformation matrix
$\bm{Q}$ contain the $n_s$ eigenvectors, which have eigenvalues $\{
\lambda \}$.  Conservation of overall population guarantees that one
of the eigenvalues $\lambda_1$ is zero, with its corresponding eigenvector $\bm{c}_{eq}$
being the equilibrium populations.  All other eigenvalues are
non-zero and negative (i.e. $\lambda_n > 0$ for $n > 1$).  Thus we can
write
\begin{eqnarray}
\bm{c}_\alpha (t) - \bm{c}_{eq} &=& \sum_{n=2}^{n_s} \bm{Q}_{\alpha
  n} \, e^{-\lambda_n t}
\, \bm{Q}^{-1}_{n \gamma} \, \bm{c}_{\gamma}(0) .
\label{matrixEvolution1}
\end{eqnarray}

When the states are ordered so that $\alpha = 1$ corresponds to 
the state with no bonds and $\alpha = 2^{n_b} = n_s$ corresponds to the state with the
maximum number of bonds (hereafter referred to as the ``folded
state''), the equilibration of the folded state
$c_f(t) = c_{n_s}(t)$ starting from an initial ensemble of non-bonded states
$\bm{c} = (1, 0, \dots ,0)$ can be monitored by tracking
\begin{eqnarray*}
\hat{c}_f(t) &=& c_f(t) - c_{f,eq} = \sum_{n=2}^{n_s} \bm{Q}_{n_s, n} \; e^{-\lambda_n
  t} \; \bm{Q}^{-1}_{n 1} \\
&=& \sum_{n=2}^{n_s} f_n e^{-\lambda_n t},
\end{eqnarray*}
where $f_n = \bm{Q}_{n_s n} \bm{Q}^{-1}_{n 1}$.  Note that 
$\hat{c}_{f}(0) = -c_{f,eq}$.
\begin{figure}[tb]
\includegraphics[width=0.6\columnwidth]{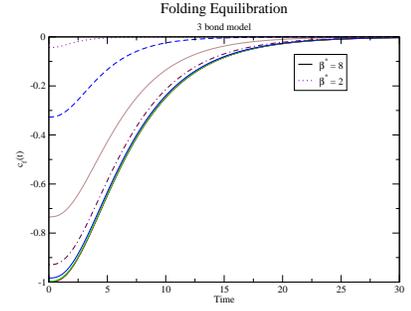}
\caption{The equilibration dynamics of the folding state population
  $\hat{c}_f(t)$ versus time for a chain of 15 monomers for values of
  inverse effective temperature $\beta^*$ ranging from $\beta^* = 2$
  (top curve, dotted lines) to $\beta^* = 8$ (solid black line, bottom
curve).}
\label{3bond}
\end{figure}

The computation of the all the inverse rates $k_1^{-1} = \langle \tau (a) \rangle$ and
$k_{-2}^{-1} = \langle \tau (b) \rangle$ necessary to compute the matrix of transition rates
$\bm{K}$ becomes tedious for model systems with many bonds
as the number of matrix elements roughly scales as the square of the
number of states $n_s$, which, in turn, scales exponentially with the number of bonds.  Fortunately,
it is not difficult to automate a procedure to analyze the data output from a series of Monte Carlo
simulations.  For the results presented below, a parallel tempering Monte Carlo (MC) sampling method coupling 
MC chains at different inverse temperature was utilized to find the relative entropies of relevant 
configurations\cite{BayatVanZonSchofield12}.
In addition, the distances of all bonding and repulsive interactions were recorded for all configurations in the
MC chain of states.  Relevant states can be identified by using a bitmap representation of configurations that translates
each configuration into a unique number based on setting bits if a bond exists in an ordered list of bonds. 
Dynamically connected states then correspond to bitmaps that differ by only a single bit.  The cumulative distribution
functions were then constructed for all dynamically connected states using a fitting procedure\cite{VanZonSchofield10}
from the data by accumulating distances for all bonds in a configuration as well
as the distances of new bonds that can be formed that lead to a connected state.  The inverse rates for
bond formation and loss were then computed by numerical integration of Eqs.~(\ref{fpa}) and (\ref{fpb})
 with a convergence factor of
$\epsilon = 1 \times 10^{-6}$.  The rate constants and rate matrix
can then be computed at any temperature.

The relaxation profile $\hat{c}_f(t)$ of a $15$-monomer chain with a maximum of $n_b
= 3$
attractive bonds is shown in Fig.~(\ref{3bond}).  For this system,
there are only $n_s = 8$ possible states, and hence $\hat{c}_f
(t)$ may be written as a superposition of $7$ exponentials.  At low
temperatures (large $\beta^*$), the dynamics becomes essentially
independent of temperature as the forward rate constants $k_f$ become
small and $k_r \approx k_{-2}$ and hence independent of temperature (assuming a
constant diffusion coefficient $D$).  At low temperatures, the first
non-zero eigenvalues are nearly triply degenerate with a value of
$\lambda_{2,3,4} \approx 0.2$, while the next set of three eigenvalues are
roughly twice as large.  Thus the dynamics is effectively single
exponential, reflecting the equivalence of three relaxation channels
in which the fully folded state $\text{BF FJ JN}$ can be reached from any
of the three precursor states with two bonds.  At low temperatures,
the dynamics essentially consists of the system falling down a series of
steps from the fully unbonded state to the fully folded (bonded) state
with little back-reaction (i.e. climbing back up the steps), and there
are multiple stairways leading to the folded state.

\begin{figure}[tb]
\includegraphics[width=0.6\columnwidth]{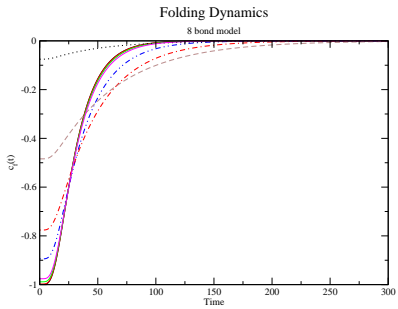}
\caption{The equilibration dynamics of the folding state population
  $\hat{c}_f(t)$ versus time for a chain of 25 monomers for values of
  inverse effective temperature $\beta^*$ ranging from $\beta^* = 2$
  (top curve, dotted lines) to $\beta^* = 8$ (solid black line, bottom
curve).}
\label{8bond}
\end{figure}

The dynamics of longer chains is much more interesting due to the
possibility of forming bonds between distant residues that mimics
the folding of secondary structural elements into tertiary
structures.  In Fig.~(\ref{8bond}), the relaxation profile of a
$25$-monomer chain is plotted versus time for a range of inverse temperatures.
The dynamics of the larger system, with $25$ monomers, $n_b=8$ attractive bonds and
$n_s = 256$ states, is much more complex than the smaller system
since $\hat{c}_f (t)$ is now a combination of $255$ exponentials and
can acquire characteristics of a stretched exponential frequently
observed in systems with frustration.  Note that the folding
time of the $25$ monomer system is typically longer than that observed in the shorter
chain, even at high temperatures.  Although there are $256$ possible states,
the nature of the interactions either geometrically prohibits many states or
makes them so improbable that they are never populated.  For the $25$ monomer system,
only $52$ states are populated and even fewer are connected through transitions,
since only transitions between states that differ by one bond are dynamically coupled in the
model.

\begin{figure}[bt]
\includegraphics[width=0.8\columnwidth]{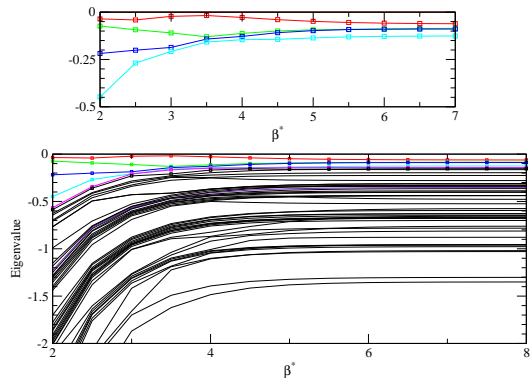}
\caption{Eigenvalues of the transition matrix of the 25 monomer system as a function
of inverse temperature.  In the top graph, the lowest $10$ eigenvalues are shown for
clarity.}
\label{matrixEigenvalues}
\end{figure}

The $51$ non-zero eigenvalues of the transition matrix for the $25$ monomer chain are shown
in Fig.~\ref{matrixEigenvalues} as a function of inverse temperature.
At high temperatures, the smallest
nonzero eigenvalue is doubly degenerate, with $\lambda_{2,3}^{(25)}
\approx 0.01$, two orders of magnitude smaller than the value of
roughly $2.0$ observed in the $15$-monomer chain.  Note that the eigenvalues
are grouped into several bands of modes due to the similarity of transitions between
states with a similar number of bonds.  The contribution of each of the eigenmodes to
the overall relaxation profile is difficult to ascertain by inspection because the expansion
coefficients $f_n$ that determine the contribution of individual modes to the relaxation profile 
can be either positive or negative and tend to cancel one another within bands.  Nonetheless, 
it is evident in Fig.~\ref{8bond} that as the temperature
decreases, the relaxation profile becomes more complex as more eigenmodes
contribute to the relaxation at longer time scales, leading to a
stretched-exponential appearance.  The folding time clearly increases
as the temperature is lowered at intermediate values of the
temperature $1 \leq \beta^* \leq 4$.  However below this temperature
regime the equilibration profile simplifies to a characterstic
single exponential form with a shorter overall folding time.  Once
again, at low effective temperatures $\beta^* \geq 6$, the relaxation
profile becomes independent of temperature and roughly single
exponential. 

The behavior of the profile $\hat{c}_f (t)$ as the temperature is
lowered can be understood in terms of the number of relaxation modes or ``folding
pathways'' that contribute to the evolution of the state
populations.  At intermediate temperatures, many modes are connected
to one another since the forward rate constants $k_f$ describing the
rate of escape from a bonding well are large enough to allow rapid
formation and loss of bonds as the system equilibrates.  However as the
temperature is lowered, the forward rate constants become small and
the relaxation proceeds as a sequence of steps of ``falling'' down the
steps in the free energy landscape.  Once again, at low temperatures
we find that $k_f \approx 0$ and $k_r \approx k_{-2}$, leading to
temperature independent dynamics.  
\section{Summary and conclusions}

In this article the dynamics of a Markov state model derived from
a protein-like chain system in which the monomers interact through
discontinuous potentials was examined.
Monte Carlo methods were utilized to compute the free energy landscape
of the discontinuous potential model as well as the potential of mean force for
the interconversion of connected configurations.  Microscopic expressions
for transitions between configurations were obtained for a chain immersed in an inert 
effective bath by analyzing
the microscopic expressions for transition rates in terms of the 
spectral decomposition of the 
projected Smoluchowski evolution operator.  This approach can be interpreted in terms of
the first-passage time solution
of the system in a two-step reaction model.  The validity of the solution was demonstrated by direct simulation
of a one-dimensional system.  

Using this apparatus, a first-order kinetic description of the relaxation of 
non-equilibrium populations was obtained in terms of a matrix of rate constants.  
The linear system was solved to characterize the relaxation profile of 
an ensemble of unfolded configurations to the folded state as a function of 
temperature.  For short chains, the relaxation is primarily single-exponential 
and becomes independent of temperature in the low-temperature regime.  The profile 
is more complicated for longer chains, where long-lived, complicated relaxation 
behavior is seen at intermediate temperatures followed by a low temperature regime 
in which the folding becomes rapid and single exponential.  The temperature dependence
of the relaxation profile was interpreted in terms of the relative contribution
of low-frequency modes that determine the evolution of populations. 

At intermediate temperatures, many modes of similar frequency 
have appreciable amplitude and thereby
contribute to the relaxation.  This regime is characterized by large forward rate constants,
describing the rate of process of escape from a bonding well.  However as the temperature is 
lowered, the forward rate constants become small and the relaxation proceeds 
as a sequence of steps of “falling” down the steps in the free energy landscape. 
As a result, temperature independent dynamics are observed at low temperatures.

The approach presented here is ideally suited for small model systems with relatively
few connected states.  Since the number of states $n_s$ increases exponentially with the
number $n_b$ of attractive interactions that form bonds, the approach becomes cumbersome
for large systems.  This unfavorable scaling with $n_b$ is mitigated somewhat since many
states have large relative free energies and are unpopulated at most temperatures.  Such
states become kinetically isolated since transitions to such states are extremely
slow.  These isolated states do not contribute meaningfully to relaxation pathways.
Furthermore, the matrix of rate constants is relatively sparse since only states that differ
by one bond are connected to one another.

The overall shape of the free energy surface for the models considered here is relatively simple
since intermediate unfolded states are smoothly connected to folded states along direct pathways.
This topology is effectively maintained over the whole temperature range, leading to dynamics that
is free of long-lived meta-stable configurations and frustration effects. Nonetheless, 
the procedure presented here can be applied to other model
systems with more complicated interactions that lead to  ``mis-folded" states provided
the relative entropy and cumulative densities of dynamically connected states can be estimated 
with adequate accuracy.  The model and methods allow the connections between structural
motifs, the topology of the free energy surface and the temperature dependence of the dynamical profile
of the folding process to be examined.

\section*{Acknowledgments}
Computations were performed on the GPC supercomputer at the SciNet HPC Consortium, 
which is funded by the Canada Foundation for Innovation under the auspices of 
Compute Canada, the Government of Ontario, the Ontario Research Fund Research Excellence 
and the University of Toronto.

This work was supported by a grant from the Natural Sciences and Engineering
Research Council of Canada.

\vspace{.1in}
\noindent
\section*{Appendix A: Reduction of the transition matrix}
In this appendix the microscopic expression for the transition rate matrix $k_{ab}(t)$ in Eq.~(\ref{NonMarkovRates})
is simplified to a low-dimensional form that is amenable to spectral analysis.
In Eq.~(\ref{NonMarkovRates}) for the transition rate matrix element $k_{ab}(t)$,
\begin{eqnarray}
k_{ab}(t) &=&  \left\langle \left( e^{\Omd \Qd t}
    \Omd \chi_a \right) \chi_b
  \right\rangle .
\label{NonMarkovRates_ab}
\end{eqnarray}
We assume that the states {\it A} and {\it B} differ by
only a single bond dependent on distance $x_k$, which implies that the indicator functions $\chi_a$ and $\chi_b$ for states
{\it A} and {\it B} obey
\begin{eqnarray*}
\chi_a &=& H_a (x_k - x_c) \chi_b^k(\rn) \nonumber \\
\chi_b &=& H_b(x_k - x_c) \chi_b^k (\rn), 
\label{states}
\end{eqnarray*}
where $\chi^k_b (\rn )$ is the indicator function for
configuration {\it B} {\it without} the factor of $H(x_k - x_c)$ or
$H^b(x_k - x_c)$ indicating whether or not $x_k \geq x_c$.
In Eq.~(\ref{states}), $H_\alpha$ is either $H^b$ or $H$ depending on whether state $\alpha$ has a $k$ bond or not.

When the trajectories only cross a single dividing surface where
$x_k = x_c$, it follows that $e^{\Q \Om t} ( \chi_b \rho_e )
= \chi_{b}^k e^{\Q_2 \Om t} ( H_{b} \rho_e )$.
The projection operator $\Q_2$ is $\Q_2 = 1 - \P_2$, where $\P_2$ 
projects only onto the indicator function for the two configurations, {\it A} and {\it B},
\begin{eqnarray*}
\P_2 f &=& \chi_b^k \rho_e \left[ \frac{H}{\langle H \rangle_k} \int_k d\rn d\pn \; H f +
\frac{H^b}{\langle H^b \rangle_k} \int_k d\rn d\pn \; H^b f \right] \nonumber \\
&=& \chi_b^k \rho_e \Pd_2 f,
\end{eqnarray*}
where $\langle g \rangle_k = \int d\rn d\pn \; \chi_b^k \rho_e g = \int_k d\rn d\pn \; \rho_e g$ denotes the 
conditional ensemble average of $g$ over the phase space for configurations which satisfy the bonding state
condition $\chi_b^k = 1$.

Under these circumstances and noting that $\Om H^b = - \Om H$, Eq.~(\ref{NonMarkovRates_ab}) can be written as
\begin{eqnarray}
k_{ab}(t) &=& \left( 2 \delta_{a,b} - 1 \right) \int d\rn d\pn \; H \chi_b^k  \Om e^{\Q_2 \Om t} \bigg( \chi_b^k 
H \rho_e \bigg) \nonumber \\
&=& \left( 2 \delta_{a,b} - 1 \right) \int_k d\rn d\pn \; H \Om \rho_b^{+} (\rn, \pn, t) ,
\label{Gamma1}
\end{eqnarray}
where $\rho_b^{+} (\rn ,\pn ,t) = e^{\Q_2 \Om t} \bigg( \chi_b^k 
H \rho_e \bigg) = e^{\Q_2 \Om t} \rho_b (\rn, \pn, 0)$ evolves under the projected dynamics.
We now reduce the dimensionality of the operator $\Om$ by focusing on the reduced probability density
$W^{+}_b(\Rs, \Ps, t)$ for the pair of phase space coordinates 
$\Rs = \{ \rn_m , \rn_n\}$ and $\Ps = \{ \pn_m, \pn_n \}$ defined
by integrating the other internal ``bath" degrees of freedom $\x_i = \{ \ri, \psubi \} = \{ \rn_k, \pn_k : k \neq m,n \}$ over the restricted
configurational phase space of the polymer
\begin{eqnarray}
W_b^{+}(\Rs, \Ps ,t) &=& \int_k d\x_i \; \rho_b^{+} (\rn, \pn ,t).
\end{eqnarray}
From Eq.~(\ref{Gamma1}), we see that
\begin{eqnarray*}
k_{ab}(t) &=& \left( 2 \delta_{a,b} - 1 \right) \int d\Rs d\Ps \; H (\Rs) \int_k d\x_i \; \Om \rho_b^{+}(t).
\end{eqnarray*}
To obtain an equation of motion for the reduced density $W^{+}_b$, we define new projection operators
$\Pb$ and $\Qb = 1 - \Pb$, where $\Pb$ operates on an arbitrary density $g(\rn, \pn)$ as
\begin{eqnarray*}
\Pb g (\rn, \pn) = \tilde{\rho}_b \int_k d\x_i \; g(\Rs, \Ps, \x_i),
\end{eqnarray*}
where
\begin{eqnarray*}
\tilde{\rho}_b &=& \frac{\chi_b^k \rho_e}{\int_k d\x_i \; \rho_e} = \frac{\chi_b^k \rho_e}{W_{e,b}}.
\end{eqnarray*}
Using the decomposition of the pair potential in Eq.~(\ref{potDecomposition}), we write the Hamiltonian
as $\H = \H_0 + \phi(\Rs, \ri ) + \H_i$ and note that
\begin{eqnarray*}
\tilde{\rho}_b &=& \chi_b^k \, \rho_i e^{-\beta ({\phi - \omega})} \nonumber \\
\rho_i &=& \frac{e^{-\beta \H_i}}{\int_k d\x_i \; e^{-\beta \H_i}} \nonumber \\
e^{-\beta \omega (\Rs)} &=& \int_k d\x_i \; \rho_i e^{-\beta \phi} ,
\end{eqnarray*}
where $\omega$ is the potential of mean force for the restricted configuration defined by $\chi_b^k = 1$.
The full density is then decomposed as $\rho_b^{+}(t) = (\Pb + \Qb) \rho_b^{+}(t) = \tilde{\rho}_b W_b^{+}(t) + z(t)$,
and straightforward application of operator identities gives for $t \gg \tau_l$\cite{Schofield93,McBride90}
\begin{eqnarray}
W_b^{+}(t) &=& e^{\Q_2 \Om_2 t} W_b^{+}(0) \nonumber \\
z(t) &=& \int_0^t d\tau \; e^{\Qb \Q_2 \Om \tau} \Q_2 {\tilde{\rho}_b} \nabla_{\Rs} 
(\phi - \omega) \cdot \nonumber \\
&&\left(\nabla_{\Ps} + \beta \frac{\Ps}{m} \right) W_b^{+}(t - \tau) ,
\label{results}
\end{eqnarray}
where the two-particle Fokker-Planck operator $\Om_2$ is given by
\begin{eqnarray}
\Om_2 &=& -\frac{\Ps}{m}\cdot \nabla_{\Rs} + \nabla_{\Rs}\left( U_0 + \omega  \right) \cdot \nabla_{\Ps} \nonumber \\
&& \; \; +
\nabla_{\Ps} \cdot \tilde{\gamma} \cdot \left( \nabla_{\Ps} + \beta \frac{\Ps}{m} \right) .
\label{FokkerPlanck2}
\end{eqnarray}
In Eq.~(\ref{FokkerPlanck2}), $\tilde{\gamma}$ is the renormalized friction matrix, which is given to leading
order in $\tau_l/\tau_c$ by
\begin{eqnarray}
\tilde{\gamma} &=& \gamma \I + \int_0^\infty d\tau \left\langle \hat{\F}  e^{\Omd_i \tau} \hat{\F} \right\rangle_b ,
\label{renormFriction}
\end{eqnarray}
where $\left\langle f \right\rangle_b = \int_k d\x_i \; f \tilde{\rho}_b $.  In Eq.~(\ref{renormFriction}),
$\F = -\nabla_{\Rs} \phi$ is the force on the bonding monomers arising from interactions with the rest of the chain,
$\hat{\F} = \F - \langle \F \rangle_b$, and 
\begin{eqnarray*}
\Om^{\dagger}_i &=& \frac{\psubi}{m} \cdot \nabla_{\ri} - \nabla_{\ri} ( U_i + \phi ) \cdot \nabla_{\psubi}
\nonumber \\
&& \; \; \; + \gamma \left( \nabla_{\psubi} - \beta 
\frac{\psubi}{m} \right) \cdot \nabla_{\psubi} 
\end{eqnarray*}
corresponds to the evolution operator of the bath degrees of freedom in the presence of fixed bonding coordinates
$\Rs$.  Note that the conditional density $\tilde{\rho}_b$ is stationary under the Hermitian conjugate of 
the operator $\Om^\dagger_i$ since $\Om_i \tilde{\rho}_b = 0$.  The second term in Eq.~(\ref{renormFriction})
corresponds to an ``internal" friction arising from the influence of all monomers on the bonding monomers.
In principle, this correlation function must be computed using either numerical  methods or kinetic theory.
Although event-driven dynamical methods can be used to 
estimate the internal friction in a similar way to the evaluation of the friction in hard sphere fluids\cite{Piasecki94},
the computation using kinetic theory of the renormalized friction for a system with discontinuous 
interactions is difficult since,
unlike simple hard sphere interactions, the infinite confining square-well potentials can lead to rapid 
recollision events.  In the simplest scenario, the motion of a confined adjacent pair of monomers can be likened
to the Langevin dynamics of a particle moving inside a reflecting sphere in which repeated, rapid glancing collisions at the
boundary of the sphere play an important role.  The magnitude of the effective friction depends on the local interactions of the bonding monomers with their neighbors and therefore depends on the local connectivity 
of bath monomers to each of the bond-forming monomers for conformations satisfying $\chi_b^k = 1$.
One expects that bonding monomers centrally-located in the chain to experience a greater effective friction
than bonding monomers located near the ends due to the greater number of local interactions through which a
binding monomer drags its sterically-connected neighbors. 
In a similar context, the intra-chain diffusion coefficient plays an important role in diffusion-controlled 
intra-chain reactions such as loop closing or intra-chain catalytic processes in the Rouse, Rouse-Zimm, and other models of polymer dynamics\cite{Fixman74b,Makarov10,Eaton08,Natz12,Bhuyan13}.

Using the results in Eq.~(\ref{results}), we find that to leading order in $\tau_l/\tau_c$, 
\begin{eqnarray*}
\int_k d\x_i \; \Om \rho^{+}_b(t) &=& \int_k d\x_i \; \Om \left[ \tilde{\rho}_b W^{+}_b(t) + z(t) \right] \nonumber \\
&=& \Om_2 W^{+}_b (t),
\end{eqnarray*}
and hence
\begin{eqnarray*}
k_{ab}(t) &=& \left( 2 \delta_{a,b} - 1 \right) \int d\Rs \; H(\Rs) \int d\Ps \; \Om_2 W_b^{+}(t) .
\end{eqnarray*}

In the high-friction limit a separation of time scale between the relaxation of momentum and spatial degrees
of freedom is anticipated.  In this ``diffusive regime", the time dependence of the reduced coordinate space 
probability density $P_b^{+} (\Rs, t) = \int d\Ps \; W_b^{+}(\Rs, \Ps, t)$ is governed by the Smoluchowski 
equation\cite{Schofield93,McBride90,Schofield96}
\begin{eqnarray}
\frac{\partial P_b^{+}(t)}{\partial t} &=& \bar{\Q} \Li P^{+}(t) + O(\delta^3),
\label{Smoluchowski2}
\end{eqnarray}
where 
\begin{eqnarray*}
\Li &=& (k_B T)^2 \nabla_{\Rs} \cdot \mathbf{\tilde{\gamma}}^{-1} \cdot \left( \nabla_{\Rs} + \beta \nabla_{\Rs}(U_0 +\omega) \right) \nonumber \\
\bar{\P} f(\Rs) &=& P_{e,b}(\Rs) \bigg( \frac{H}{\langle H \rangle} \int d\Rs \; H(\Rs) f(\Rs) \nonumber \\
&& + \frac{H^b}{\langle H^b \rangle}
\int d\Rs \; H^b(\Rs) f(\Rs) \bigg) \nonumber \\
&=& P_{e,b}(\Rs) \bar{\P^\dagger} f(\Rs) \nonumber \\
\bar{\Q} &=& 1 - \bar{\P},
\end{eqnarray*}
with $P_{e,b}(\Rs) = \int d\Ps \; W_{e,b}(\Rs, \Ps)$.  In Eq.~(\ref{Smoluchowski2}), the small parameter $\delta$ is associated with the smooth spatial derivatives of $P_b^{+}(\Rs,t)$.
Once again projecting out the fast degrees of freedom yields
\begin{eqnarray*}
\int d\Ps \; \Om_2 W_b^{+}(t) = \Li P_b^{+} (t),
\end{eqnarray*}
so that
\begin{eqnarray*}
k_{ab}(t) &=& \left( 2 \delta_{a,b} - 1 \right) \int d\Rs \; H \Li P_b^{+}(t),
\end{eqnarray*}
where $P_b^{+}(t) = \exp\{\bar{\Q}\Li t\} \left( H P_{e,b} \right)$.
Therefore in the diffusive regime,
the elements of the rate matrix may be expressed as
\begin{eqnarray*}
K_{ab}(t) &=& \frac{ 2\delta_{a,b} -1}{\langle \chi_b \rangle}
 \int d\Rs \; P_{e,b} \left( e^{\Li^\dagger \bar{\Q}^\dagger t} \Li^\dagger H \right) H .
\end{eqnarray*}

In what follows, we will assume that the generalized friction matrix $\tilde{\gamma} = \tilde{\gamma}\I$ 
is diagonal and independent
of $\Rs$, so the Smoluchowski operator simplifies to
\begin{eqnarray*}
\Li &=& \tilde{D} \nabla_{\Rs} \cdot \left( \nabla_{\Rs} + \beta \nabla_{\Rs}(U_0 +\omega) \right) ,
\end{eqnarray*}
where the effective diffusion constant is given by $\tilde{D} = (k_BT)^2/\tilde{\gamma}$.  Subsequently, we will also
neglect any differences arising in the diffusion coefficient that are due to the specific location of the bonding
pair $m$ and $n$ in the chain since the effect of the location of the bonding pair within the chain on the diffusion coefficient is not known. 

Since the projected dynamics is assumed to involve only one change of bond in
the rate constants $K_{ab}$, the projected Smoluchowski
operator $\Li^\dagger \bar{\Q}$ may be further simplified when operating on a function
$g(x)$ of the scalar distance $x=r_{mn}$ between a pair of monomers $m$ and $n$ by defining center of mass
and relative coordinates so that
\begin{eqnarray*}
\Li^\dagger \bar{\Q}^\dagger g(x) &=&  \Lid_{x} {\Q}_x^\dagger \, g(x) = \Lid_{xq} \, g(x) \nonumber \\
&=& D \left( \frac{d^2}{dx^2} - \beta {\phi}_{ab}^\prime (x) \frac{d}{dx} \right) {\Q}_x^\dagger g(x),
\end{eqnarray*}
where the prime notation indicates the derivative,
$D = 2\tilde{D}$ and the potential of mean force $\phi_{ab}(x) = - k_B T \, \log\rho_{ab}(x)$, with
$\rho_{ab}(x) = \langle \delta (x - r_{mn})
\chi_{b}^k (\rn ) \rangle = \int d\Rs \; \delta (x - r_{mn}) P_{e,b}(\Rs)$.  
Finally, we have the useful result that for $a \neq b$,
\begin{equation*}
k_{ab}(t) =  \int_{x_\text{min}}^{x_\text{max}} dx \; \rho_{ab}(x) 
\, H(x - x_c)  \, e^{\Lid_{xq} t} \Lid_x H(x-x_c) , 
\label{k_alphabeta}
\end{equation*}
which is Eq.~(\ref{k_alphabeta1}) in the main text.

\section*{Appendix B: Eigenvalue analysis of rate constants}
In this appendix we obtain explicit expressions for the forward and backward rate constants by identifying the eigenvector $\xi_1(x)$ with zero eigenvalue.
We first discuss the effect of the discontinuities in the potential of mean force on the boundary conditions
for the eigenvectors of the Smoluchowski equation and then evaluate the first eigenvectors.

Considering the effect of the discontinuities on the eigenvalue equations, 
recall that the potential of mean force and its 
derivative have a finite jump from $\phi(x_c^-)$ to $\phi(x_c^+)$ and $\phi^\prime(x_c^-)$
to $\phi^\prime(x_c^+)$, respectively.  The singular behavior may be expressed as jump
conditions at the singularities for the eigenvectors by noting that the Smoluchowski equation for the 
one-dimensional probability
density $W(x,t)$ should have a continuous probability current $S(x,t)$ 
at all times $t$ and points $x$\cite{Vollmer79}. The Smoluchowski equation 
for the one-dimensional system is given by
\begin{eqnarray}
\frac{\partial W(x,t)}{\partial t} &=& \Li_x W(x,t) = - \frac{\partial S(x,t)}{\partial x}, \label{fp}
\end{eqnarray}
where the unprojected Smoluchowski operator is 
\begin{eqnarray*}
\Li_x = D \frac{\partial}{\partial x} \left( \frac{\partial}{\partial x} + \beta \phi^\prime (x) \right).
\end{eqnarray*}
Since the probability current must be continuous at $x=x_c$, $S(x_c^+,t) = S(x_c^{-},t)$, which implies
\begin{equation}
\begin{split}
\beta \phi^\prime (x_c^+) &W(x_c^+,t) + \frac{\partial W}{\partial x}\bigg|_{x=x_c^+}  \\
&=  \beta \phi^\prime (x_c^-) W(x_c^-,t) + \frac{\partial W}{\partial x}\bigg|_{x=x_c^-}
\end{split}
 \label{Jump1}
\end{equation}
and
\begin{eqnarray}
e^{\beta \phi (x_c^+)} W(x_c^+,t) = e^{\beta \phi (x_c^-)} W(x_c^-,t).
\label{Jump2}
\end{eqnarray}
These conditions can be expressed as jump conditions for the eigenvectors $\xi_n(x)$ by expanding the probability density
in terms of the complete set of eigenvectors of the unprojected operator $\Li_x$,
\begin{eqnarray*}
W(x,t) = \sum_{n=0}^{\infty} W_n q_n(x) e^{-\tilde{\lambda}_n t},
\end{eqnarray*}
where $\Li_x q_n (x) = -{\tilde{\lambda}_n} q_n(x)$.  Noting
that Eqs.~(\ref{Jump1}) and (\ref{Jump2}) hold for arbitrary times,
we obtain
\begin{eqnarray}
e^{\beta \phi (x_c^+)} q_n (x_c^+) &=& e^{\beta \phi (x_c^-)} q_n (x_c^-) \label{jump1} \\
\beta \phi^\prime (x_c^+) q_n (x_c^+) + q_n^\prime (x_c^+) 
&=&  \beta \phi^\prime (x_c^-) q_n (x_c^-) \nonumber \\
&& \; \; + q_n^\prime (x_c^-). \label{jump2} 
\end{eqnarray}
Since we can write eigenvalues $\xi_n(x)$ of the projected evolution operator $\Li_{xq}$ as a position-independent linear combination of the eigenvectors $q_n(x)$,
\begin{eqnarray*}
\xi_m (x) = \sum_{n=0}^{\infty} \alpha_{mn} q_n(x),
\end{eqnarray*} 
where the coefficients $\alpha_{mn}$ are constant,
Eqs.~(\ref{jump1}) and (\ref{jump2}) hold identically for
the $\xi_n (x)$ as well.  A discussion of the relationships between the eigenvectors $q_n(x)$ and $\xi_n (x)$ can
be found in Ref.~[22].

To solve the eigenvalue equations in Eq.~(\ref{eigenvalueEqn}),
we first note that since $\Li_x \rho(x) = 0$, $\Q_x \Li_x \rho (x) = 0$ and
hence $\xi_0(x) = \rho (x)$, $\mu_0 = 0$  and $\xid_0(x) = 1$.  For the unprojected
dynamics, this is the only eigenvector with zero eigenvalue.  However there is
another eigenvector $\xi_1(x)$ of the operator $\Q_x \Li_x$ with zero eigenvalue 
that arises due to the operator $\Q_x$.   

The second eigenvector with zero eigenvalue of the operator $\Lid_x \Qd_x$ is easy to
obtain.  Noting that $\Qd_x H(x-x_c) = \Qd_x H^b(x -x_c) = 0$, we can write
$\Lid_x \Qd_x \xid_1 (x) = 0$ with
\begin{eqnarray*}
\xid_1(x) &=& \alpha_1 H(x-x_c) + \alpha_2 H^b(x-x_c).
\end{eqnarray*}
From the orthogonality condition $\langle \xi_1 | \xi_0 \rangle = 0$, we find that 
\begin{eqnarray*}
\xid_1(x) = b_1 \left( \frac{H(x-x_c)}{\langle H(x-x_c) \rangle} - \frac{H^b(x
  -x_c)}{\langle H^b(x - x_c) \rangle} \right),
\end{eqnarray*}
where $b_1 = \langle \xi_1 |H(x-x_c) \rho (x) \rangle$ is constant.  
Hence $\xid_1(x)$ is an eigenvector of the $\Lid_{xq}$ operator with zero
eigenvalue, $\mu_1^\dagger = 0$. 

To find  $\xi_1(x)$, we note that 
\begin{eqnarray*}
\Q_x \Li \xi_1(x) &=& (1 - \P_x) \Li \xi_1 (x) = 0 \\
\Li \xi_1 (x) &=& \P_x \Li \xi_1 (x),
\end{eqnarray*}
which requires that for $x \geq x_c$, 
\begin{eqnarray}
\Li \xi_1 (x) &=& D \frac{d}{dx} \left( \frac{d}{dx} + \beta \phi^\prime_b
  (x) \right) \xi_1(x) \nonumber \\
&=& a_1 H(x-x_c) \rho_a (x) \label{a1}, 
\end{eqnarray}
and for $x < x_c$, 
\begin{eqnarray}
\Li \xi_1 (x) &=& D \frac{d}{dx} \left( \frac{d}{dx} + \beta \phi^\prime_a
  (x) \right) \xi_1(x) \nonumber \\
&=& -a_1 (1-H(x - x_c)) \rho_a (x) \nonumber \\
&=& -a_1 H^b(x-x_c) \rho_a (x). \label{a2}
\end{eqnarray}

Integrating Eq.~(\ref{a1}) from $x_c$ to $y$ yields
\begin{eqnarray*}
\left( \frac{d}{dy} + \beta \phi^\prime_b
  (y) \right) \xi_1(y) &=& -a_1 \big( 1-C_b(y) \big) ,
\end{eqnarray*}
where $C_b(y)$ is the cumulative distribution function $C_b(y) = \int_{x_c}^{y} dx \, \rho_b(x)$.
Noting that 
\begin{eqnarray*}
\left( \frac{d}{dy} + \beta \phi^\prime \right)\xi_1(y) = e^{-\beta \phi(y)}
\frac{d}{dy} \left( e^{\beta \phi(y)} \xi_1(y) \right) ,
\end{eqnarray*}
and integrating over $y$ from $x_c$ to $x$ for $x > x_c$ gives, 
\begin{equation}
\begin{split}
\xi_1(x) &= C_1 e^{-\beta \phi_b (x)}  \\
&- e^{-\beta \phi_b (x)} \frac{a_1}{D} \int_{x_c}^{x} dy \; e^{\beta \phi_b (y)} \big( 1 - C_b(y) \big)  ,
\end{split}
\label{eigenPart1}
\end{equation}
where $C_1 = \exp\{\beta \phi_b(x_c^{+})\} \xi_1(x_c^+)$ is a constant.
Similarly, using the same approach that led to
Eq.~(\ref{eigenPart1}), we find from Eq.~(\ref{a2}) that for $x < x_c$,
\begin{equation*}
\xi_1(x) = \tilde{C}_1 e^{-\beta \phi_b(x)}  
+ \frac{a_1}{D} e^{-\beta \phi_b(x)}
\int_{x}^{x_c} dy \; e^{\beta \phi_a (y)} C_a(y) ,
\end{equation*}
where $\tilde{C}_1 = \exp\{\beta \phi_b(x_c^{-})\} \xi_1(x_c^-)$.  
The jump discontinuity condition in Eq.~(\ref{jump1}) implies
\begin{eqnarray*}
e^{\beta \phi_b (x_c^+)} \xi_1(x_c^+) = C_1 =  e^{\beta \phi_a (x_c^-)} \xi_1 (x_c^-) = \tilde{C}_1 , 
\end{eqnarray*}
so we get
\begin{eqnarray*}
\xi_1 (x) &=& C_1 \left( H(x-x_c) e^{-\beta \phi_b(x)}  + H^b(x-x_c) e^{-\beta \phi_a(x)} \right) \nonumber \\
&&+ \frac{a_1}{D} J(x)  \nonumber 
\end{eqnarray*}
where 
\begin{eqnarray*}
J(x) &=& H^b(x - x_c) \rho_a(x) \int_{x}^{x_c} dy \;  \frac{C_a(y)}{\rho_a(y)}\nonumber \\
  &&- H(x-x_c) \rho_b(x) \int_{x_c}^{x} dy \; \frac{\big( 1 - C_b(y) \big)}{\rho_b(y)} .
\end{eqnarray*}
Since $\langle \xi_0 | \xi_1 \rangle = 0 = \int_{x_\text{min}}^{x_\text{max}} dx \; \xi_1 (x)$, we must have
\begin{eqnarray*}
C_1 = - \frac{1}{Z_A + Z_B} \frac{a_1}{D} \int_{x_\text{min}}^{x_\text{max}} dx \; J(x),
\end{eqnarray*}
leading to the general solution
\begin{eqnarray*}
\xi_1(x) = \frac{a_1}{D} \bigg[ \left(  -\int_{x_\text{min}}^{x_\text{max}} dy \, J(y) \right)
  \rho(x) +  J(x) \bigg].
\label{eigen}
\end{eqnarray*}
From the orthonormality condition $\langle \xi_1 | \xi_1 \rangle = 1$,
we find that since $\langle \xi_1 | \xi_0 \rangle = 0$, 
\begin{eqnarray*}
\frac{D}{-a_1 b_1} &=& \frac{-1}{b_1} \int_{x_\text{min}}^{x_\text{max}} \xi_1^\dagger (x) J(x) \nonumber \\
&=& \frac{1}{\langle
  H^b(x -x_c) \rangle} \int_{x_\text{min}}^{x_c} dx \; \frac{C_a(x)^2}{\rho_a(x)} \nonumber \\
  &&+ 
\frac{1}{\langle
  H(x -x_c) \rangle} \int_{x_c}^{x_\text{max}} dx \; \frac{(1-C_b(x))^2}{\rho_b(x)} ,
\end{eqnarray*}
leading to Eq.~(\ref{finalForwardBackward}).

\bibliographystyle{apsrev}

\begin{thebibliography}{48}
\expandafter\ifx\csname natexlab\endcsname\relax\def\natexlab#1{#1}\fi
\expandafter\ifx\csname bibnamefont\endcsname\relax
  \def\bibnamefont#1{#1}\fi
\expandafter\ifx\csname bibfnamefont\endcsname\relax
  \def\bibfnamefont#1{#1}\fi
\expandafter\ifx\csname citenamefont\endcsname\relax
  \def\citenamefont#1{#1}\fi
\expandafter\ifx\csname url\endcsname\relax
  \def\url#1{\texttt{#1}}\fi
\expandafter\ifx\csname urlprefix\endcsname\relax\def\urlprefix{URL }\fi
\providecommand{\bibinfo}[2]{#2}
\providecommand{\eprint}[2][]{\url{#2}}

\bibitem[{\citenamefont{Wales}(2003)}]{WalesBook}
\bibinfo{author}{\bibfnamefont{D.}~\bibnamefont{Wales}},
  \emph{\bibinfo{title}{Energy Landscapes}} (\bibinfo{publisher}{Cambridge
  University Press}, \bibinfo{address}{Cambridge}, \bibinfo{year}{2003}).

\bibitem[{\citenamefont{No\'e et~al.}(2007)\citenamefont{No\'e, Sch\"utte, and
  Smith}}]{Noe07}
\bibinfo{author}{\bibfnamefont{I.}~\bibnamefont{No\'e},
  \bibfnamefont{F.and~Horenko}},
  \bibinfo{author}{\bibfnamefont{C.}~\bibnamefont{Sch\"utte}},
  \bibnamefont{and} \bibinfo{author}{\bibfnamefont{J.}~\bibnamefont{Smith}},
  \bibinfo{journal}{J. Chem. Phys.} \textbf{\bibinfo{volume}{126}},
  \bibinfo{pages}{155102} (\bibinfo{year}{2007}).

\bibitem[{\citenamefont{Buchete and Hummer}(2008)}]{Hummer08}
\bibinfo{author}{\bibfnamefont{N.-V.} \bibnamefont{Buchete}} \bibnamefont{and}
  \bibinfo{author}{\bibfnamefont{G.}~\bibnamefont{Hummer}},
  \bibinfo{journal}{J. Phys. Chem. B} \textbf{\bibinfo{volume}{112}},
  \bibinfo{pages}{6057} (\bibinfo{year}{2008}).

\bibitem[{\citenamefont{Karpen et~al.}(1993)\citenamefont{Karpen, Tobias, and
  Brooks}}]{Karpen93}
\bibinfo{author}{\bibfnamefont{M.}~\bibnamefont{Karpen}},
  \bibinfo{author}{\bibfnamefont{D.}~\bibnamefont{Tobias}}, \bibnamefont{and}
  \bibinfo{author}{\bibfnamefont{C.}~\bibnamefont{Brooks}},
  \bibinfo{journal}{Biochemistry} \textbf{\bibinfo{volume}{32}},
  \bibinfo{pages}{412} (\bibinfo{year}{1993}).

\bibitem[{\citenamefont{Prinz et~al.}(2011{\natexlab{a}})\citenamefont{Prinz,
  Chodera, Pande, Swope, Smith, and No\'e}}]{Prinz11}
\bibinfo{author}{\bibfnamefont{J.-H.} \bibnamefont{Prinz}},
  \bibinfo{author}{\bibfnamefont{J.}~\bibnamefont{Chodera}},
  \bibinfo{author}{\bibfnamefont{V.}~\bibnamefont{Pande}},
  \bibinfo{author}{\bibfnamefont{W.}~\bibnamefont{Swope}},
  \bibinfo{author}{\bibfnamefont{J.}~\bibnamefont{Smith}}, \bibnamefont{and}
  \bibinfo{author}{\bibfnamefont{F.}~\bibnamefont{No\'e}}, \bibinfo{journal}{J.
  Chem. Phys.} \textbf{\bibinfo{volume}{134}}, \bibinfo{pages}{244108}
  (\bibinfo{year}{2011}{\natexlab{a}}).

\bibitem[{\citenamefont{Chodera et~al.}(2007)\citenamefont{Chodera, Dill,
  Singhal, Pande, Swope, and Pitera}}]{Chodera07}
\bibinfo{author}{\bibfnamefont{J.}~\bibnamefont{Chodera}},
  \bibinfo{author}{\bibfnamefont{K.}~\bibnamefont{Dill}},
  \bibinfo{author}{\bibfnamefont{N.}~\bibnamefont{Singhal}},
  \bibinfo{author}{\bibfnamefont{V.}~\bibnamefont{Pande}},
  \bibinfo{author}{\bibfnamefont{W.}~\bibnamefont{Swope}}, \bibnamefont{and}
  \bibinfo{author}{\bibfnamefont{J.}~\bibnamefont{Pitera}},
  \bibinfo{journal}{J. Chem. Phys.} \textbf{\bibinfo{volume}{126}},
  \bibinfo{pages}{155101} (\bibinfo{year}{2007}).

\bibitem[{\citenamefont{Pan and Roux}(2008)}]{Pan08}
\bibinfo{author}{\bibfnamefont{A.}~\bibnamefont{Pan}} \bibnamefont{and}
  \bibinfo{author}{\bibfnamefont{B.}~\bibnamefont{Roux}}, \bibinfo{journal}{J.
  Chem. Phys.} \textbf{\bibinfo{volume}{129}}, \bibinfo{pages}{064107}
  (\bibinfo{year}{2008}).

\bibitem[{\citenamefont{Caves et~al.}(1998)\citenamefont{Caves, Evanseck, and
  Karplus}}]{Caves98}
\bibinfo{author}{\bibfnamefont{L.}~\bibnamefont{Caves}},
  \bibinfo{author}{\bibfnamefont{J.}~\bibnamefont{Evanseck}}, \bibnamefont{and}
  \bibinfo{author}{\bibfnamefont{M.}~\bibnamefont{Karplus}},
  \bibinfo{journal}{Protein Sci.} \textbf{\bibinfo{volume}{7}},
  \bibinfo{pages}{649} (\bibinfo{year}{1998}).

\bibitem[{\citenamefont{Chodera et~al.}(2006)\citenamefont{Chodera, Swope,
  Pitera, and Dill}}]{Chodera06}
\bibinfo{author}{\bibfnamefont{J.}~\bibnamefont{Chodera}},
  \bibinfo{author}{\bibfnamefont{W.}~\bibnamefont{Swope}},
  \bibinfo{author}{\bibfnamefont{J.}~\bibnamefont{Pitera}}, \bibnamefont{and}
  \bibinfo{author}{\bibfnamefont{K.}~\bibnamefont{Dill}},
  \bibinfo{journal}{Multiscale Model. Simul.} \textbf{\bibinfo{volume}{5}},
  \bibinfo{pages}{1214} (\bibinfo{year}{2006}).

\bibitem[{\citenamefont{Bowman et~al.}(2009)\citenamefont{Bowman, Beauchamp,
  Boxer, and Pande}}]{Bowman09}
\bibinfo{author}{\bibfnamefont{G.}~\bibnamefont{Bowman}},
  \bibinfo{author}{\bibfnamefont{K.}~\bibnamefont{Beauchamp}},
  \bibinfo{author}{\bibfnamefont{G.}~\bibnamefont{Boxer}}, \bibnamefont{and}
  \bibinfo{author}{\bibfnamefont{V.}~\bibnamefont{Pande}}, \bibinfo{journal}{J.
  Chem. Phys.} \textbf{\bibinfo{volume}{131}}, \bibinfo{pages}{124101}
  (\bibinfo{year}{2009}).

\bibitem[{\citenamefont{Keller et~al.}(2010)\citenamefont{Keller, Daura, and
  van Gunsteren}}]{Keller10}
\bibinfo{author}{\bibfnamefont{B.}~\bibnamefont{Keller}},
  \bibinfo{author}{\bibfnamefont{X.}~\bibnamefont{Daura}}, \bibnamefont{and}
  \bibinfo{author}{\bibfnamefont{W.}~\bibnamefont{van Gunsteren}},
  \bibinfo{journal}{J. Chem. Phys.} \textbf{\bibinfo{volume}{132}},
  \bibinfo{pages}{074110} (\bibinfo{year}{2010}).

\bibitem[{\citenamefont{Prinz et~al.}(2011{\natexlab{b}})\citenamefont{Prinz,
  Wu, Sarich, Keller, Senne, Held, Chodera, Sch\"utte, and No\'e}}]{Prinz11a}
\bibinfo{author}{\bibfnamefont{J.-H.} \bibnamefont{Prinz}},
  \bibinfo{author}{\bibfnamefont{H.}~\bibnamefont{Wu}},
  \bibinfo{author}{\bibfnamefont{M.}~\bibnamefont{Sarich}},
  \bibinfo{author}{\bibfnamefont{B.}~\bibnamefont{Keller}},
  \bibinfo{author}{\bibfnamefont{M.}~\bibnamefont{Senne}},
  \bibinfo{author}{\bibfnamefont{M.}~\bibnamefont{Held}},
  \bibinfo{author}{\bibfnamefont{J.}~\bibnamefont{Chodera}},
  \bibinfo{author}{\bibfnamefont{C.}~\bibnamefont{Sch\"utte}},
  \bibnamefont{and} \bibinfo{author}{\bibfnamefont{F.}~\bibnamefont{No\'e}},
  \bibinfo{journal}{J. Chem. Phys.} \textbf{\bibinfo{volume}{134}},
  \bibinfo{pages}{174105} (\bibinfo{year}{2011}{\natexlab{b}}).

\bibitem[{\citenamefont{Tirion}(1996)}]{Tirion96}
\bibinfo{author}{\bibfnamefont{M.}~\bibnamefont{Tirion}},
  \bibinfo{journal}{Phys. Rev. Lett.} \textbf{\bibinfo{volume}{77}},
  \bibinfo{pages}{1905} (\bibinfo{year}{1996}).

\bibitem[{\citenamefont{Smith and Hall}(2001)}]{Hall2001a}
\bibinfo{author}{\bibfnamefont{A.}~\bibnamefont{Smith}} \bibnamefont{and}
  \bibinfo{author}{\bibfnamefont{C.}~\bibnamefont{Hall}},
  \bibinfo{journal}{Proteins} \textbf{\bibinfo{volume}{44}},
  \bibinfo{pages}{344} (\bibinfo{year}{2001}).

\bibitem[{\citenamefont{Shirvanyants et~al.}(2012)\citenamefont{Shirvanyants,
  Ding, Tsao, Ramachandran, and Dokholyan}}]{Dokholyan2012}
\bibinfo{author}{\bibfnamefont{D.}~\bibnamefont{Shirvanyants}},
  \bibinfo{author}{\bibfnamefont{F.}~\bibnamefont{Ding}},
  \bibinfo{author}{\bibfnamefont{D.}~\bibnamefont{Tsao}},
  \bibinfo{author}{\bibfnamefont{S.}~\bibnamefont{Ramachandran}},
  \bibnamefont{and}
  \bibinfo{author}{\bibfnamefont{N.}~\bibnamefont{Dokholyan}},
  \bibinfo{journal}{J. Phys. Chem. B} \textbf{\bibinfo{volume}{116}},
  \bibinfo{pages}{8375} (\bibinfo{year}{2012}).

\bibitem[{\citenamefont{Ding et~al.}(2008)\citenamefont{Ding, Tsao, Nie, and
  Dokholyan}}]{Ding2008}
\bibinfo{author}{\bibfnamefont{F.}~\bibnamefont{Ding}},
  \bibinfo{author}{\bibfnamefont{D.}~\bibnamefont{Tsao}},
  \bibinfo{author}{\bibfnamefont{H.}~\bibnamefont{Nie}}, \bibnamefont{and}
  \bibinfo{author}{\bibfnamefont{N.}~\bibnamefont{Dokholyan}},
  \bibinfo{journal}{Structure} \textbf{\bibinfo{volume}{16}},
  \bibinfo{pages}{1010} (\bibinfo{year}{2008}).

\bibitem[{\citenamefont{Nguyen and Hall}(2004{\natexlab{a}})}]{Hall2004a}
\bibinfo{author}{\bibfnamefont{H.}~\bibnamefont{Nguyen}} \bibnamefont{and}
  \bibinfo{author}{\bibfnamefont{C.}~\bibnamefont{Hall}},
  \bibinfo{journal}{Proc. Natl. Acad. Sci. USA} \textbf{\bibinfo{volume}{101}},
  \bibinfo{pages}{16180} (\bibinfo{year}{2004}{\natexlab{a}}).

\bibitem[{\citenamefont{Nguyen and Hall}(2004{\natexlab{b}})}]{Hall2004b}
\bibinfo{author}{\bibfnamefont{H.}~\bibnamefont{Nguyen}} \bibnamefont{and}
  \bibinfo{author}{\bibfnamefont{C.}~\bibnamefont{Hall}},
  \bibinfo{journal}{Biophys. J.} \textbf{\bibinfo{volume}{87}},
  \bibinfo{pages}{4122} (\bibinfo{year}{2004}{\natexlab{b}}).

\bibitem[{\citenamefont{Urbanc et~al.}(2006)\citenamefont{Urbanc, Borreguero,
  Cruz, and Stanley}}]{Stanley2006}
\bibinfo{author}{\bibfnamefont{B.}~\bibnamefont{Urbanc}},
  \bibinfo{author}{\bibfnamefont{J.~M.} \bibnamefont{Borreguero}},
  \bibinfo{author}{\bibfnamefont{L.}~\bibnamefont{Cruz}}, \bibnamefont{and}
  \bibinfo{author}{\bibfnamefont{H.~E.} \bibnamefont{Stanley}},
  \bibinfo{journal}{Methods in Enzymology} \textbf{\bibinfo{volume}{412}},
  \bibinfo{pages}{314} (\bibinfo{year}{2006}).

\bibitem[{\citenamefont{Bayat-Movahed et~al.}(2012)\citenamefont{Bayat-Movahed,
  van Zon, and Schofield}}]{BayatVanZonSchofield12}
\bibinfo{author}{\bibfnamefont{H.}~\bibnamefont{Bayat-Movahed}},
  \bibinfo{author}{\bibfnamefont{R.}~\bibnamefont{van Zon}}, \bibnamefont{and}
  \bibinfo{author}{\bibfnamefont{J.}~\bibnamefont{Schofield}},
  \bibinfo{journal}{J. Chem. Phys.} \textbf{\bibinfo{volume}{136}},
  \bibinfo{pages}{245103} (\bibinfo{year}{2012}).

\bibitem[{\citenamefont{Bellemans et~al.}(1980)\citenamefont{Bellemans, Orban,
  and van Belle}}]{Bellemans:26}
\bibinfo{author}{\bibfnamefont{A.}~\bibnamefont{Bellemans}},
  \bibinfo{author}{\bibfnamefont{J.}~\bibnamefont{Orban}}, \bibnamefont{and}
  \bibinfo{author}{\bibfnamefont{D.}~\bibnamefont{van Belle}},
  \bibinfo{journal}{Mol. Phys.} \textbf{\bibinfo{volume}{39}},
  \bibinfo{pages}{781} (\bibinfo{year}{1980}).

\bibitem[{\citenamefont{Zhou and Karplus}(1997)}]{ZHOU:15}
\bibinfo{author}{\bibfnamefont{Y.}~\bibnamefont{Zhou}} \bibnamefont{and}
  \bibinfo{author}{\bibfnamefont{M.}~\bibnamefont{Karplus}},
  \bibinfo{journal}{Proc. Natl. Acad. Sci. USA} \textbf{\bibinfo{volume}{94}},
  \bibinfo{pages}{14429} (\bibinfo{year}{1997}).

\bibitem[{\citenamefont{van Zon and Schofield}(2010)}]{VanZonSchofield10}
\bibinfo{author}{\bibfnamefont{R.}~\bibnamefont{van Zon}} \bibnamefont{and}
  \bibinfo{author}{\bibfnamefont{J.}~\bibnamefont{Schofield}},
  \bibinfo{journal}{J. Chem. Phys.} \textbf{\bibinfo{volume}{132}},
  \bibinfo{pages}{154110} (\bibinfo{year}{2010}).

\bibitem[{\citenamefont{Mazur and Oppenheim}(1970)}]{Oppenheim70}
\bibinfo{author}{\bibfnamefont{P.}~\bibnamefont{Mazur}} \bibnamefont{and}
  \bibinfo{author}{\bibfnamefont{I.}~\bibnamefont{Oppenheim}},
  \bibinfo{journal}{Physica} \textbf{\bibinfo{volume}{50}},
  \bibinfo{pages}{241} (\bibinfo{year}{1970}).

\bibitem[{\citenamefont{Wu and Kapral}(1989)}]{Kapral89}
\bibinfo{author}{\bibfnamefont{X.-G.} \bibnamefont{Wu}} \bibnamefont{and}
  \bibinfo{author}{\bibfnamefont{R.}~\bibnamefont{Kapral}},
  \bibinfo{journal}{J. Chem. Phys.} \textbf{\bibinfo{volume}{91}},
  \bibinfo{pages}{5528} (\bibinfo{year}{1989}).

\bibitem[{\citenamefont{Morita}(1994)}]{Morita94}
\bibinfo{author}{\bibfnamefont{A.}~\bibnamefont{Morita}},
  \bibinfo{journal}{Phys. Rev. E} \textbf{\bibinfo{volume}{49}},
  \bibinfo{pages}{3697} (\bibinfo{year}{1994}).

\bibitem[{\citenamefont{Morita}(1996)}]{Morita96}
\bibinfo{author}{\bibfnamefont{A.}~\bibnamefont{Morita}}, \bibinfo{journal}{J.
  Mol. Liquids} \textbf{\bibinfo{volume}{69}}, \bibinfo{pages}{31}
  (\bibinfo{year}{1996}).

\bibitem[{\citenamefont{Felderhof}(2008)}]{Felderhof08}
\bibinfo{author}{\bibfnamefont{B.}~\bibnamefont{Felderhof}},
  \bibinfo{journal}{Physica A} \textbf{\bibinfo{volume}{387}},
  \bibinfo{pages}{39} (\bibinfo{year}{2008}).

\bibitem[{\citenamefont{Kramers}(1940)}]{Kramers40}
\bibinfo{author}{\bibfnamefont{H.}~\bibnamefont{Kramers}},
  \bibinfo{journal}{Physica} \textbf{\bibinfo{volume}{7}}, \bibinfo{pages}{284}
  (\bibinfo{year}{1940}).

\bibitem[{\citenamefont{Szabo et~al.}(1980)\citenamefont{Szabo, Schulten, and
  Schulten}}]{Schulten80}
\bibinfo{author}{\bibfnamefont{A.}~\bibnamefont{Szabo}},
  \bibinfo{author}{\bibfnamefont{K.}~\bibnamefont{Schulten}}, \bibnamefont{and}
  \bibinfo{author}{\bibfnamefont{Z.}~\bibnamefont{Schulten}},
  \bibinfo{journal}{J. Chem. Phys.} \textbf{\bibinfo{volume}{72}},
  \bibinfo{pages}{4350} (\bibinfo{year}{1980}).

\bibitem[{\citenamefont{Schulten et~al.}(1981)\citenamefont{Schulten, Schulten,
  and Szabo}}]{Schulten81}
\bibinfo{author}{\bibfnamefont{K.}~\bibnamefont{Schulten}},
  \bibinfo{author}{\bibfnamefont{Z.}~\bibnamefont{Schulten}}, \bibnamefont{and}
  \bibinfo{author}{\bibfnamefont{A.}~\bibnamefont{Szabo}}, \bibinfo{journal}{J.
  Chem. Phys.} \textbf{\bibinfo{volume}{74}}, \bibinfo{pages}{4426}
  (\bibinfo{year}{1981}).

\bibitem[{\citenamefont{Malevanets and Kapral}(1999)}]{MalevanetsKapral99}
\bibinfo{author}{\bibfnamefont{A.}~\bibnamefont{Malevanets}} \bibnamefont{and}
  \bibinfo{author}{\bibfnamefont{R.}~\bibnamefont{Kapral}},
  \bibinfo{journal}{J. Chem. Phys.} \textbf{\bibinfo{volume}{110}},
  \bibinfo{pages}{8605} (\bibinfo{year}{1999}).

\bibitem[{\citenamefont{Malevanets and Kapral}(2000)}]{MalevanetsKapral00}
\bibinfo{author}{\bibfnamefont{A.}~\bibnamefont{Malevanets}} \bibnamefont{and}
  \bibinfo{author}{\bibfnamefont{R.}~\bibnamefont{Kapral}},
  \bibinfo{journal}{J. Chem. Phys.} \textbf{\bibinfo{volume}{112}},
  \bibinfo{pages}{7260} (\bibinfo{year}{2000}).

\bibitem[{\citenamefont{Schofield et~al.}(2012)\citenamefont{Schofield, Inder,
  and Kapral}}]{SchofieldInderKapral12}
\bibinfo{author}{\bibfnamefont{J.}~\bibnamefont{Schofield}},
  \bibinfo{author}{\bibfnamefont{P.}~\bibnamefont{Inder}}, \bibnamefont{and}
  \bibinfo{author}{\bibfnamefont{R.}~\bibnamefont{Kapral}},
  \bibinfo{journal}{J. Chem. Phys.} \textbf{\bibinfo{volume}{136}},
  \bibinfo{pages}{205101} (\bibinfo{year}{2012}).

\bibitem[{\citenamefont{Gompper et~al.}(2009)\citenamefont{Gompper, Ihle,
  Kroll, and Winkler}}]{GIKW09}
\bibinfo{author}{\bibfnamefont{G.}~\bibnamefont{Gompper}},
  \bibinfo{author}{\bibfnamefont{T.}~\bibnamefont{Ihle}},
  \bibinfo{author}{\bibfnamefont{D.~M.} \bibnamefont{Kroll}}, \bibnamefont{and}
  \bibinfo{author}{\bibfnamefont{R.~G.} \bibnamefont{Winkler}},
  \bibinfo{journal}{{A}dv. {P}olym. {S}ci.} \textbf{\bibinfo{volume}{221}},
  \bibinfo{pages}{1} (\bibinfo{year}{2009}).

\bibitem[{\citenamefont{Kikuchi et~al.}(2002)\citenamefont{Kikuchi, Gent, and
  Yeomans}}]{kikuchi02}
\bibinfo{author}{\bibfnamefont{N.}~\bibnamefont{Kikuchi}},
  \bibinfo{author}{\bibfnamefont{A.}~\bibnamefont{Gent}}, \bibnamefont{and}
  \bibinfo{author}{\bibfnamefont{J.~M.} \bibnamefont{Yeomans}},
  \bibinfo{journal}{Eur. Phys. J.} \textbf{\bibinfo{volume}{9}},
  \bibinfo{pages}{63} (\bibinfo{year}{2002}).

\bibitem[{\citenamefont{Kikuchi et~al.}(2003)\citenamefont{Kikuchi, Pooley,
  Ryder, and Yeomans}}]{kikuchi03}
\bibinfo{author}{\bibfnamefont{N.}~\bibnamefont{Kikuchi}},
  \bibinfo{author}{\bibfnamefont{C.}~\bibnamefont{Pooley}},
  \bibinfo{author}{\bibfnamefont{J.}~\bibnamefont{Ryder}}, \bibnamefont{and}
  \bibinfo{author}{\bibfnamefont{J.~M.} \bibnamefont{Yeomans}},
  \bibinfo{journal}{J. Chem. Phys.} \textbf{\bibinfo{volume}{119}},
  \bibinfo{pages}{6388} (\bibinfo{year}{2003}).

\bibitem[{\citenamefont{Stehfest}(1970)}]{Stehfest70a}
\bibinfo{author}{\bibfnamefont{H.}~\bibnamefont{Stehfest}},
  \bibinfo{journal}{Comm. ACM} \textbf{\bibinfo{volume}{13}},
  \bibinfo{pages}{47} (\bibinfo{year}{1970}).

\bibitem[{\citenamefont{Schofield and Oppenheim}(1993)}]{Schofield93}
\bibinfo{author}{\bibfnamefont{J.}~\bibnamefont{Schofield}} \bibnamefont{and}
  \bibinfo{author}{\bibfnamefont{I.}~\bibnamefont{Oppenheim}},
  \bibinfo{journal}{Physica A} \textbf{\bibinfo{volume}{196}},
  \bibinfo{pages}{209} (\bibinfo{year}{1993}).

\bibitem[{\citenamefont{Oppenheim and McBride}(1990)}]{McBride90}
\bibinfo{author}{\bibfnamefont{I.}~\bibnamefont{Oppenheim}} \bibnamefont{and}
  \bibinfo{author}{\bibfnamefont{J.}~\bibnamefont{McBride}},
  \bibinfo{journal}{Physica A} \textbf{\bibinfo{volume}{165}},
  \bibinfo{pages}{279} (\bibinfo{year}{1990}).

\bibitem[{\citenamefont{Bocquet et~al.}(1994)\citenamefont{Bocquet, Hansen, and
  Piasecki}}]{Piasecki94}
\bibinfo{author}{\bibfnamefont{L.}~\bibnamefont{Bocquet}},
  \bibinfo{author}{\bibfnamefont{J.-P.} \bibnamefont{Hansen}},
  \bibnamefont{and} \bibinfo{author}{\bibfnamefont{J.}~\bibnamefont{Piasecki}},
  \bibinfo{journal}{J. Stat. Phys.} \textbf{\bibinfo{volume}{76}},
  \bibinfo{pages}{527} (\bibinfo{year}{1994}).

\bibitem[{\citenamefont{Wilemski and Fixman}(1974)}]{Fixman74b}
\bibinfo{author}{\bibfnamefont{G.}~\bibnamefont{Wilemski}} \bibnamefont{and}
  \bibinfo{author}{\bibfnamefont{M.}~\bibnamefont{Fixman}},
  \bibinfo{journal}{J. Chem. Phys.} \textbf{\bibinfo{volume}{60}},
  \bibinfo{pages}{878} (\bibinfo{year}{1974}).

\bibitem[{\citenamefont{Makarov}(2010)}]{Makarov10}
\bibinfo{author}{\bibfnamefont{D.}~\bibnamefont{Makarov}}, \bibinfo{journal}{J.
  Chem. Phys.} \textbf{\bibinfo{volume}{132}}, \bibinfo{pages}{035104}
  (\bibinfo{year}{2010}).

\bibitem[{\citenamefont{Cellmer et~al.}(2008)\citenamefont{Cellmer, Henry,
  Hofrichter, and Eaton}}]{Eaton08}
\bibinfo{author}{\bibfnamefont{T.}~\bibnamefont{Cellmer}},
  \bibinfo{author}{\bibfnamefont{E.}~\bibnamefont{Henry}},
  \bibinfo{author}{\bibfnamefont{J.}~\bibnamefont{Hofrichter}},
  \bibnamefont{and} \bibinfo{author}{\bibfnamefont{W.}~\bibnamefont{Eaton}},
  \bibinfo{journal}{Proc. Natl. Acad. Sci.} \textbf{\bibinfo{volume}{105}},
  \bibinfo{pages}{18320} (\bibinfo{year}{2008}).

\bibitem[{\citenamefont{Schulz et~al.}(2012)\citenamefont{Schulz, Schmidt,
  Best, Dzubiella, and Netz}}]{Natz12}
\bibinfo{author}{\bibfnamefont{J.}~\bibnamefont{Schulz}},
  \bibinfo{author}{\bibfnamefont{L.}~\bibnamefont{Schmidt}},
  \bibinfo{author}{\bibfnamefont{R.}~\bibnamefont{Best}},
  \bibinfo{author}{\bibfnamefont{J.}~\bibnamefont{Dzubiella}},
  \bibnamefont{and} \bibinfo{author}{\bibfnamefont{R.}~\bibnamefont{Netz}},
  \bibinfo{journal}{J. Am. Chem. Soc.} \textbf{\bibinfo{volume}{134}},
  \bibinfo{pages}{6273} (\bibinfo{year}{2012}).

\bibitem[{\citenamefont{Yasin et~al.}(2013)\citenamefont{Yasin, Sashi, and
  Bhuyan}}]{Bhuyan13}
\bibinfo{author}{\bibfnamefont{U.}~\bibnamefont{Yasin}},
  \bibinfo{author}{\bibfnamefont{P.}~\bibnamefont{Sashi}}, \bibnamefont{and}
  \bibinfo{author}{\bibfnamefont{A.}~\bibnamefont{Bhuyan}},
  \bibinfo{journal}{J. Phys. Chem. B} \textbf{\bibinfo{volume}{117}},
  \bibinfo{pages}{12059} (\bibinfo{year}{2013}).

\bibitem[{\citenamefont{Schofield et~al.}(1996)\citenamefont{Schofield, Marcus,
  and Rice}}]{Schofield96}
\bibinfo{author}{\bibfnamefont{J.}~\bibnamefont{Schofield}},
  \bibinfo{author}{\bibfnamefont{A.}~\bibnamefont{Marcus}}, \bibnamefont{and}
  \bibinfo{author}{\bibfnamefont{S.}~\bibnamefont{Rice}}, \bibinfo{journal}{J.
  Phys. Chem.} \textbf{\bibinfo{volume}{100}}, \bibinfo{pages}{18950}
  (\bibinfo{year}{1996}).

\bibitem[{\citenamefont{M\"orsch et~al.}(1979)\citenamefont{M\"orsch, Risken,
  and Vollmer}}]{Vollmer79}
\bibinfo{author}{\bibfnamefont{M.}~\bibnamefont{M\"orsch}},
  \bibinfo{author}{\bibfnamefont{H.}~\bibnamefont{Risken}}, \bibnamefont{and}
  \bibinfo{author}{\bibfnamefont{H.~D.} \bibnamefont{Vollmer}},
  \bibinfo{journal}{Z. Phys. B} \textbf{\bibinfo{volume}{32}},
  \bibinfo{pages}{245} (\bibinfo{year}{1979}).

\end{thebibliography}

\end{document}